\documentclass[preprint,showpacs,amsmath,amssymb,prd]{revtex4-1}
\usepackage{epsf}
\usepackage{amsmath,amsthm,amssymb}
\usepackage{color}
\usepackage{dcolumn}
\usepackage{bm}
\usepackage{graphicx,epsfig}
\usepackage[dvipsnames]{xcolor}

\graphicspath{{fig/}}
\everymath{\displaystyle}

\begin{document} 

\title{Position and spin in relativistic quantum mechanics}

\author{Liping Zou$^{1}$}
\email{zoulp@impcas.ac.cn}
\author{Pengming Zhang$^{2}$}
\email{zhangpm5@mail.sysu.edu.cn}
\author{Alexander J. Silenko$^{1,3,4}$}
\email{alsilenko@mail.ru}

\affiliation{$^1$Institute of Modern Physics, Chinese Academy of
Sciences, Lanzhou 730000, China}

\affiliation{$^2$School of Physics and Astronomy, Sun Yat-sen University,
Zhuhai 519082, China}

\affiliation{$^3$Bogoliubov Laboratory of Theoretical Physics, Joint
Institute for Nuclear Research, Dubna 141980, Russia}

\affiliation{$^4$Research Institute for Nuclear Problems, Belarusian
State University, Minsk 220030, Belarus}

\date{\today}

\begin{abstract}
The problem of the position and spin in relativistic quantum mechanics
is analyzed in detail. It is definitively shown that the position and
spin operators in the Foldy-Wouthuysen representation (but not in the
Dirac one) are quantum-mechanical counterparts of the classical position
and spin variables. The probabilistic interpretation is valid only for
Foldy-Wouthuysen wave functions. The relativistic spin operators are
discussed. The spin-orbit interaction does not exist for a free particle
if the conventional operators of the orbital angular momentum and the
rest-frame spin are used. Alternative definitions of the orbital angular
momentum and the spin are based on noncommutative geometry, do not
satisfy standard commutation relations, and can allow the spin-orbit interaction.
\end{abstract}

\maketitle

\section{Introduction}

The position operator is very important for relativistic quantum
mechanics (QM). In nonrelativistic Schr\"{o}dinger QM, this operator
is equal to the radius vector $\bm r$. However, a transition to
relativistic QM leads to a dependence of this operator on a
representation. It has been shown by Pryce \cite{Pryce} that the
form of the position operator for a spin-1/2 particle is nontrivial
and some possible forms have been obtained. Newton and Wigner
\cite{NewtonWigner} have obtained the form of the
position operator having commuting components and localized
eigenfunctions in the manifold of positive-energy
wave functions based on the Dirac representation. Foldy and Wouthuysen
have shown \cite{FW} that this operator is equal to the radius vector
operator in the Foldy-Wouthuysen (FW) representation.

The spin angular momentum (or the spin for short) takes one of
central places in relativistic QM.
The spin of a Dirac (spin-1/2) particle
is defined by the $2\times2$ Pauli
matrices $\sigma_i~(i=1,2,3)$ which generate together with the unit
matrix an irreducible representation of the SU(2) group. The
Pauli matrices are Hermitian, unitary, and traceless. The classical
spin is connected with the three-dimensional rotation group SO(3).
Algebraically, SU(2) is the double covering group of SO(3).
This relation plays an important role in the
theory of rotations of spinors in nonrelativistic QM. As a result,
the spin dynamics defined by the Schr\"{o}dinger-Pauli equation
fully corresponds to the classical picture of rotation of the spin
in external fields. When quadrupole and other multipole interactions
are neglected, the spin dynamics of particles with higher spins
$(s>1/2)$ is very similar to that of a spin-1/2 particle. In particular,
the angular velocity of spin rotation depends on the electric and magnetic
dipole moments of a particle and does not explicitly depend on the spin
quantum number $s$.
The spin operator of a nonrelativistic spin-1/2 particle,
$\bm s=\hbar\bm\sigma/2$, fully corresponds to the classical spin.

A clear correspondence between quantum-mechanical operators and classical
variables is a distinguishing feature of nonrelativistic QM. This
correspondence takes place for all main operators including position
(coordinate), momentum, and angular momentum ones.

In contrast, the connection between the quantum-mechanical operators
and classical variables in relativistic QM is not so simple.
It is well known that the Dirac equation distorts the connection among
the energy, momentum, and velocity operators. Nevertheless, the problem
of operators of relativistic QM corresponding to basic classical variables
has been definitively solved in the 1960s. Moreover,
the correct and definite solution of this problem
is already contained in the famous paper by Foldy and
Wouthuysen \cite{FW}. It has been established that quantum-mechanical
counterparts of the classical variables
of the radius vector (position), momentum, angular momentum, and spin
of a Dirac particle are the operators $\bm x,~\bm p,
~\bm L=\bm x\times\bm p$, and $\bm s=\hbar\bm\Sigma/2$
defined in the \emph{Foldy-Wouthuysen} (FW) representation.
These conclusions agree with the results obtained by
Pryce \cite{Pryce} and Newton and Wigner \cite{NewtonWigner} and have been
confirmed in a lot of publications.

Unfortunately, these achievements were not reflected in textbooks
and currently many researchers hold the opposite view.
After more than sixty years, the scientific literature is full of
incorrect (explicit or implicit) statements that the position and
angular momentum of a particle are defined by the operators $\bm r$
and $\bm r\times\bm p$ in the \emph{Dirac} representation. Similarly,
one often uses definitions of the spin operator different from the
operator obtained in the fundamental works by Foldy and Wouthuysen
\cite{FW} and Fradkin and Good \cite{FG}. Such definitions may lead
to a spin-orbit interaction (SOI) for a free particle.
This situation is very typical, in particular, in physics of twisted
(vortex) electrons (see the reviews
\cite{BliokhSOI,Lloyd}). The description of the particle position by
the \emph{Dirac} radius vector is so common that the papers containing
the right description \cite{BarnettPRL,ResonanceTwistedElectrons}
were followed by the Comments \cite{BBBarnett,BB}.
A short analysis of the problem has been given in Ref. \cite{Reply2019}.
In the present work, we reproduce known (but forgotten) arguments
in favor of a definite connection between classical variables and
corresponding operators which shows the special role of the FW
representation. We also put forward some arguments given by a
contemporary development of theory of the FW transformation. These
arguments relate to a description of spinning particles \emph{in external fields}. In the
present study, we focus our attention on the spin while problems
connected with the operators of the position and the angular momentum
are also properly addressed.

The paper is organized as follows. In the next section, we explain
main distinguishing features of the relativistic FW transformation.
In Sec. \ref{Classical},
we reproduce the past (but forgotten) approach to
carrying out an unambiguous determination of basic operators for a
free Dirac particle and the corresponding classical variables. This
approach leads to the definitions of fundamental operators of the
position and spin which were generally accepted sixty years ago but
have been unreasonably revised lately. Important additional arguments
based on the relativistic FW transformation in external fields are
presented in Sec. \ref{Comparison}. Section \ref{Operators} describes
the relativistic operators of the position and spin. The related
problems of relativistic QM (a probabilistic interpretation of a
wave function, spin-orbit interaction for a free particle and \textit{Zitterbewegung})
are expounded in Sec. \ref{selected}. The results are discussed and
summarized in Sec. \ref{Discussions}.

We use the system of units $\hbar=1,~c=1$. We include $\hbar$ and $c$
explicitly when this inclusion
clarifies the problem. The square and curly brackets, $[\dots,\dots]$
and $\{\dots,\dots\}$,
denote commutators and Poisson brackets, respectively. The standard
denotations of Dirac matrices are applied (see, e.g., Ref. \cite{BLP}).
In particular, $\bm\Sigma\equiv{\rm diag}(\bm\sigma,\bm\sigma),~\bm\Pi=\beta\bm\Sigma$.

\section{Relativistic Foldy-Wouthuysen transformation}\label{Relativistic}

The connection between fundamental classical variables and the
corresponding operators is studied in the framework of relativistic
QM and the FW representation happens to be very useful. Therefore,
a consideration of the relativistic FW transformation is instructive.

In this section, we focus our attention on such a transformation
for a particle in external fields. However,
a consideration of relativistic particles with different spins in external fields is
not simple because of specific properties of initial equations. All
these equations substantially differ from the Schr\"{o}dinger equation
of nonrelativistic QM. The Dirac equation in the Hamiltonian form
corrupts the connection among energy, momentum, and velocity. The
connection between the relativistic QM and Schr\"{o}dinger QM is
restored by the FW transformation \cite{FW}. In the FW representation,
the relativistic QM takes the Schr\"{o}dinger form. This fact has been
first shown \cite{FW} for a nonrelativistic Dirac particle in
electromagnetic fields and for a free relativistic Dirac particle.

The important development of QM in the FW representation has been made
in Ref. \cite{E} where the exact FW transformation operator has been
derived and main properties of this operator have been determined. This
operator is defined by
   \begin{equation}
   \Psi_{FW}=U_{FW}\Psi_D\equiv\exp{(iS_{FW})}\Psi_D,
   \label{eqUiS}
   \end{equation}
where $S_{FW}$ is the exponential FW transformation operator. The
transformation is unitary ($U_{FW}^\dag=U_{FW}^{-1}$). There
is an infinite set of representations different from the FW
representation whose distinctive feature is a block-diagonal
form of the Hamiltonian. The FW transformation is \emph{uniquely} defined
by the condition that the exponential
operator $S_{FW}$ is \emph{odd},
\begin{equation}
\beta S_{FW}=-S_{FW}\beta,
\label{VveEfrt}
\end{equation}
and Hermitian \cite{E,erik}. This condition is equivalent to \cite{E,erik}
\begin{equation}
\beta U_{FW}=U^\dag_{FW}\beta.
\label{Erikcon}
\end{equation}

Eriksen \cite{E} found the exact expression for the nonexponential
FW transformation operator. It is convenient to present this expression
in the form \cite{JMPcond}
 \begin{equation}
U_{FW}=U_{E}=\frac{1+\beta\lambda}{\sqrt{2+\beta\lambda+\lambda\beta}},
~~~ \lambda=\frac{{\cal H}}{({\cal H}^2)^{1/2}}.
\label{eqXXI}
\end{equation} To unambiguously define
the square root, these relations should be complemented by the condition that the square root of
the unit matrix $\mathcal{I}$ is equal to the unit matrix \cite{JMP}.
The exact exponential FW transformation operator has been determined
in Ref. \cite{PRAExpO}.
The initial Hamiltonian operator ${\cal H}$ is arbitrary. It
is easy to see that \cite{E}
\begin{equation}
\lambda^2=1, \quad [\beta\lambda,\lambda\beta]=0,
\quad [\beta,(\beta\lambda+\lambda\beta)]=0.
\label{eq3X3}
\end{equation}

The equivalent form of the operator $U_{E}$ \cite{JMPcond} shows
that it is properly unitary:
\begin{equation}
U_{E}=\frac{1+\beta\lambda}{\sqrt{(1+\beta\lambda)^\dag(1+\beta\lambda)}}.
\label{JMP2009}
\end{equation}
The additional substantiation of the Eriksen method was presented
in Ref. \cite{Valid}.

However, Eq. (\ref{JMP2009}) containing square roots of operators
is not applicable for a derivation of relativistic expressions for
FW Hamiltonians except for a few special cases \cite{FW,Valid,Case}.
Equation (\ref{JMP2009}) can be used for a calculation of series of
relativistic corrections to nonrelativistic FW Hamiltonians.

Many transformation methods allowing one to derive a block-diagonal
Hamiltonian do not lead to the FW representation (see Refs.
\cite{E,erik,JMPcond} for more details). Paradoxically, the original
FW method \cite{FW} does not satisfy the Eriksen conditions and does
not lead to the FW representation \cite{erik,dVFor}. Any FW
transformation method satisfying these conditions is correct. All
FW Hamiltonians obtained by correct methods \emph{coincide}. Methods
which do not satisfy the Eriksen conditions can be corrected. For the
original FW method \cite{FW}, such corrections have been obtained in
Refs. \cite{erik,PRAExpO,dVFor,TMPFW,PRA2016}.

Contemporary QM requires relativistic methods giving compact
relativistic FW Hamiltonians for any energy. The first such
Hamiltonian has been derived by Blount \cite{Blount}. At present,
there are many different relativistic FW transformation methods
(see Refs. \cite{electrodynamics,PRA2016,PRA2015,ChiouChen} and
references therein). In the present work, we use the results
obtained by the method proposed in Ref. \cite{JMP} and then
developed in Refs. \cite{PRAExpO,TMPFW,PRA2016,PRA2015,PRA2008}.
The validity of this method has been rigorously proven in Ref.
\cite{PRA2015}. The general form of the initial Hamiltonian is
given by \cite{PRA2008}
\begin{equation}
{\cal H}=\beta{\cal M}+{\cal E}+{\cal
O},~~~\beta{\cal M}={\cal M}\beta,
~~~\beta{\cal E}={\cal E}\beta,
~~~\beta{\cal O}=-{\cal O}\beta.
\label{eq3}
\end{equation}
The
even operators ${\cal M}$ and ${\cal E}$ and the odd operator
${\cal O}$ are diagonal and off-diagonal in two spinors,
respectively. Equation (\ref{eq3}) is applicable for a particle with any
spin if the number of components of a corresponding wave function
is equal to $2(2s+1)$, where $s$ is the spin quantum number. For a
Dirac particle, the ${\cal M}$ operator is usually equal to the
particle mass $m$:
\begin{equation}
{\cal H}_D=\beta m+{\cal E}+{\cal
O}.
\label{eq3Dirac}
\end{equation}

The approximate nonexponential FW transformation operator can
be presented as follows \cite{PRA2015}:
\begin{equation}
\begin{array}{c}
U=\frac{1+\sqrt{1+X^2}+\beta X}{\sqrt{2\sqrt{1+X^2}\left(1+\sqrt{1+X^2}\right)}},
\quad X=\left\{\frac{1}{2{\cal M}},{\cal O}\right\}.
\end{array}
\label{nexptop}
\end{equation}
The approximate relativistic FW Hamiltonian is given by \cite{PRA2015}
\begin{equation}
\begin{array}{c}
{\cal H}_{FW}=\beta\epsilon+ {\cal E}+\frac 14\left\{\frac{1}
{2\epsilon^2+\{\epsilon,{\cal M}\}},\left(\beta\left[{\cal O},[{\cal O},{\cal
M}]\right]-[{\cal O},[{\cal O},{\cal
F}]]\right)\right\}, \quad \epsilon=\sqrt{{\cal M}^2+{\cal
O}^2}.
\end{array}
\label{MHamf}
\end{equation}

As an example, we can consider a spin-1/2 particle interacting
with electromagnetic fields. If the particle possesses the
anomalous magnetic moment (AMM) $\mu'$ and the electric dipole
moment (EDM) $d$, its interaction is defined by the Dirac equation
added by the Pauli term and the term proportional to the
EDM (see Ref. \cite{RPJ}):
\begin{equation}
\biggl(i\gamma^\mu
D_\mu-m+\frac{\mu'}{2}\sigma^{\mu\nu}F_{\mu\nu}+\frac{d}{2}\sigma^{\mu\nu}G_{\mu\nu}\biggr)
\Psi=0,
\label{eqDPgen}
\end{equation}
where $D_\mu=\partial_\mu+ieA_\mu$ is the covariant derivative,
$G_{\mu\nu}=(-\bm B,\bm E)$ is the tensor dual to the
electromagnetic field one, and $F_{\mu\nu}=(\bm E,\bm B)$.

The Dirac-Pauli Hamiltonian added by the EDM terms has
the form (\ref{eq3Dirac}) where ${\cal E}$ and ${\cal
O}$ are defined by
\begin{equation}
{\cal E}=e\Phi-\mu'\bm
{\Pi}\cdot \bm{B}-d\bm {\Pi}\cdot \bm{E}, \qquad {\cal
O}=\bm{\alpha}\cdot\bm{\pi}+i\mu'\bm{\gamma}\cdot\bm{
E}-id\bm{\gamma}\cdot\bm{B}.
\label{DireqEDM}
\end{equation}

The calculated FW Hamiltonian is given by \cite{RPJ}
\begin{equation}
{\cal H}_{FW}={\cal H}_{FW}^{(MDM)}+{\cal
H}_{FW}^{(EDM)},
\label{FWHamEDM}
\end{equation}
\begin{equation}
\begin{array}{c}
{\cal H}_{FW}^{(MDM)}=\beta\epsilon'+e\Phi+\frac
   14\left\{\left(\frac{\mu_0m}{\epsilon'
   +m}+\mu'\right)\frac{1}{\epsilon'},\Bigl[\bm\Sigma\cdot(\bm\pi
\times\bm E-\bm E\times\bm\pi)-\hbar\nabla
\cdot\bm E\Bigr]\right\}\\
-\frac
12\left\{\left(\frac{\mu_0m}{\epsilon'}
+\mu'\right), \bm\Pi\!\cdot\!\bm B\right\}\\
+\beta\frac{\mu'}{4}\left\{\frac{1}{\epsilon'(\epsilon'+m)},
\Bigl[(\bm{B}\!\cdot\!\bm\pi)(\bm{\Sigma}\!\cdot\!\bm\pi)+ (\bm{\Sigma}
\!\cdot\!\bm\pi)(\bm\pi\!\cdot\!\bm{B})+2\pi\hbar(\bm\pi\!\cdot\!\bm j+
\bm j\!\cdot\! \bm\pi)\Bigr]\right\},
\end{array}
\label{eq33new}
\end{equation}
\begin{equation}
\begin{array}{c}
{\cal H}_{FW}^{(EDM)}=-d\bm\Pi\!\cdot\!\bm E
+\frac{d}{4}\left\{\frac{1}{\epsilon'(\epsilon'+m)},
\biggl[(\bm{E}\!\cdot\!\bm\pi)(\bm{\Pi}\!\cdot\!\bm\pi)+ (\bm{\Pi}
\!\cdot\!\bm\pi)(\bm\pi\!\cdot\!\bm{E})\biggr]\right\} \\
-\frac
d4\left\{\frac{1}{\epsilon'},\biggl(\bm\Sigma\!\cdot\![\bm\pi\!
\times\!\bm B]-\bm\Sigma\!\cdot\![\bm
B\!\times\!\bm\pi]\biggr)\right\}.
\end{array}
\label{EDMeq12}
\end{equation}
Here ${\cal H}_{FW}^{(MDM)}$ defines the contribution from the
magnetic dipole moment (MDM), $\mu_0=e\hbar/(2m)$ is the Dirac
magnetic moment, $\epsilon'=\sqrt{m^2+\bm{\pi}^2}$, and
$$
\bm j=\frac{1}{4\pi}\left(c\,\nabla\!\times\!\bm B-\frac{\partial \bm E}
{\partial t}\right)
$$
is the density of external electric current. The term in Eq.
(\ref{eq33new}) proportional to $\nabla
\cdot\bm E$ defines the Darwin (contact) interaction. While we
take into account in Eq. (\ref{EDMeq12}) terms proportional to
$\hbar^2$ and describing contact interactions with external
charges and currents, such terms are zero due to the Maxwell equations
$$
\nabla\cdot\bm B=0, \quad \nabla\times\bm E=-\frac{\partial \bm B}
{\partial t}.
$$
Terms proportional to the second and higher
powers of $\hbar$ and quadratic and bilinear in $\bm E$ and $\bm B$
are neglected. This Hamiltonian will be used in Sec. \ref{Comparison}.

When $[{\cal E},{\cal O}]=0$, the FW transformation of the Dirac
Hamiltonian (\ref{eq3Dirac}) is exact \cite{JMP}. For a free Dirac
particle, the FW transformation operator
is given by \cite{FW}
\begin{equation}
U_{FW}=\frac{\epsilon+m+\bm\gamma\cdot\bm p}{\sqrt{2\epsilon(\epsilon+m)}},
\quad U_{FW}^{-1}=\frac{\epsilon+m-\bm\gamma\cdot\bm p}{\sqrt{2\epsilon(\epsilon+m)}},\quad
\epsilon=\sqrt{m^2+\bm p^2}.
\label{eqFWfree}
\end{equation}

\section{Position and spin operators for a free Dirac particle}
\label{Classical}

A definite connection among the position, angular momentum,
and spin operators for a free Dirac particle and the corresponding
classical variables was one of great achievements of QM in the last
century.
Unfortunately, this brilliant achievement was lately revised without
appropriate substantiations. An incorrect interpretation of these
operators is now so pervasive that it fully covers the theory of
twisted (vortex) particles and is often applied in other branches
of physics and also in quantum chemistry
(see Refs. \cite{ReiherWolfBook,electrondensity}).

In this section, we reproduce the previously well-known results
allowing an unambiguous determination of basic operators for the
free Dirac particles. We follow the approach based on Refs.
\cite{Pryce,Dirac} and developed in Refs. \cite{CurrieRevModPhys,JordanMuku}.

The theory of a dynamical system is built up in terms of a number
of algebraic quantities, called
dynamical variables, each of which is defined with respect to a
system of spacetime coordinates. There are ten independent
\emph{fundamental quantities}
$P_\mu=(H,\bm P),~ J_{\mu\nu}~(\mu,\nu=0,1,2,3)$ describing the
momentum and total angular momentum and characteristic for the
dynamical system \cite{Pryce,Dirac,CurrieRevModPhys,JordanMuku}.
The antisymmetric tensor $J_{\mu\nu}$ is defined by the two vectors,
$\bm J$ and $\bm K$. As a result, there are the ten infinitesimal
generators of the Poincar\'{e} group (inhomogeneous Lorentz group
\cite{Pryce}), namely, the generators of the infinitesimal space
translations $\bm P=(P_i)$,
the generator of the infinitesimal time translation $H$, the
generators of infinitesimal rotations $\bm J=(J_i)$, and the
generators of infinitesimal Lorentz transformations (boosts)
$\bm K=(K_i)~(i=1,2,3)$ \cite{Pryce,Dirac,CurrieRevModPhys,JordanMuku,
BakamjianThomas,Foldy56,Foldy61,Bacry88}.
These ten generators satisfy the following Poisson brackets
\cite{Pryce,Dirac,CurrieRevModPhys,JordanMuku,BakamjianThomas,Foldy56,Foldy61}:
\begin{equation}
\begin{array}{c}
\{P_i,P_j\}=0,\quad \{P_i,H\}=0,\quad \{J_i,H\}=0,\\
\{J_i,J_j\}=e_{ijk}J_k,\quad \{J_i,P_j\}=e_{ijk}P_k,\quad \{J_i,K_j\}=e_{ijk}K_k,\\
\{K_i,H\}=P_{i},\quad \{K_i,K_j\}=-e_{ijk}J_k,\quad \{K_i,P_j\}=\delta_{ij}H.
\end{array}
\label{Poisson}
\end{equation}
In the multiparticle case, the momenta and energies of particles
are additive, $\bm P=\sum_k{\bm P^{(k)}}$, $H=\sum_k{H^{(k)}}$.
Counterparts of these generators in QM are ten corresponding operators.
A connection between the classical and quantum mechanics manifests
itself in the fact that the commutators of these operators are equal
to the corresponding Poisson brackets multiplied by the imaginary
unit $i$. Equation (\ref{Poisson}) describes the Lie algebra of
classical motion for a free particle which leads to the ten-dimensional
Poincar\'{e} algebra. The \emph{only} additional equation which should
be satisfied defines the orbital and spin parts of the total angular momentum:
\begin{equation}
\bm J=\bm L+\bm S,\qquad \bm L\equiv\bm Q\times\bm P.
\label{spinOAM}
\end{equation}
There is a latitude in the definition of the position, orbital
angular momentum (OAM), and spin. An exhaustive list of appropriate
definitions has been presented in Ref. \cite{Pryce}.

A consideration of the particle position variables $Q_i$ brings the
following Poisson brackets \cite{Pryce,CurrieRevModPhys,JordanMuku}:
\begin{equation}
\begin{array}{c}
\{Q_i,P_j\}=\delta_{ij},\quad \{Q_i,J_j\}=e_{ijk}Q_k,\quad \{Q_i,K_j\}=\frac12\left(Q_j
\{Q_i,H\}+\{Q_i,H\}Q_j\right)-t\delta_{ij}.
\end{array}
\label{position}
\end{equation}
The last term in the relation for $\{Q_i,K_j\}$ has been missed in
Refs. \cite{CurrieRevModPhys,JordanMuku}.
It follows from Eqs. (\ref{Poisson})-(\ref{position}) that
\begin{equation}
\begin{array}{c}
\{L_i,P_j\}=e_{ijk}P_k,\qquad \{S_i,P_j\}=0.
\end{array}
\label{addeq}
\end{equation}

Equations (\ref{Poisson})-(\ref{addeq}) should be
satisfied for any correct definition of fundamental variables.
However, these equations do not uniquely define the fundamental
variables and different sets of the variables $\bm Q, \bm L, \bm S$
can be used \cite{Pryce}.

The \emph{conventional} particle position defines
the \emph{center of charge} of a charged particle if the particle
EDM is negligible. The term ``mass point'' is also useful. For a
single particle, the \emph{mass point} always coincides with the
conventional particle position. It also coincides with the center
of charge of a charged particle when the particle EDM is neglected.
Under this assumption, the mass point is the center of both positive
and negative charges of an uncharged particle like a neutron. The
Poisson brackets for the conventional particle position are equal to zero:
\begin{equation}
\{Q_i,Q_j\}=0.
\label{positioncommutation}
\end{equation}
The property (\ref{positioncommutation}) is equivalent to the
commutativity of the particle position operators [cf. Eq. (\ref{kcommfq})]
and is nontrivial (see Ref. \cite{Pryce,JordanMuku}). Other sets of
fundamental variables violating Eq. (\ref{positioncommutation}) can
also be used \cite{Pryce}. We will consider this problem in Secs.
\ref{Operators} and \ref{selected}.
Equations (\ref{Poisson})-(\ref{positioncommutation}) describe a
classical Hamiltonian system.

The well-known deep connection between the Poisson brackets in classical
mechanics (CM) and the commutators in QM also takes place in this case.
It is important that this connection remains valid \emph{in any representation}.
We need only to present the corresponding commutation relations for free
spinning Dirac fermions. These relations allow one to establish definite
forms of operators corresponding to basic classical variables in the
Dirac and FW representations.

In the framework of CM, Eqs. (\ref{Poisson})-(\ref{positioncommutation})
allow one to obtain the following Poisson brackets
\cite{Pryce,JordanMuku,AcharyaSudarshan}
\begin{equation}
\{Q_i,L_j\}=e_{ijk}Q_k,\quad \{Q_i,S_j\}=0,
\quad \{P_i,S_j\}=0,\quad \{L_i,L_j\}=e_{ijk}L_k,\quad \{S_i,S_j\}=e_{ijk}S_k.
\label{spinOAMope}
\end{equation}
Evidently,
\begin{equation}
\{L_i,S_j\}=0.
\label{spinOAMbrackets}
\end{equation}

The main variables of a free spinning particle in CM are specified
by Eqs. (\ref{spinOAM}) and
\begin{equation}
H=\sqrt{m^2+\bm P^2}, \qquad \bm K=\bm QH-\frac{\bm S\times\bm P}{m+H}-t\bm P
\label{HK}
\end{equation}
(see also Refs. \cite{Foldy56,Foldy61} and Eq. (A.23) in
Ref. \cite{SuttorpDeGroot}).
In Refs. \cite{JordanMuku,BakamjianThomas,SuttorpDeGroot,Bacry88},
the last term in the relation for $\bm P$ has been missed.

The Poisson brackets (\ref{spinOAMope}) and (\ref{spinOAMbrackets})
show that the variable $\bm Q$ defined by Eq.  (\ref{positioncommutation})
does not depend on the spin and is the same for
spinning and spinless particles with equal $\bm Q,\bm P$, and $H$. For a
particle ensemble, the variable $\bm Q$ defines the position of the center
of charge. Otherwise, a violation of the condition (\ref{positioncommutation})
leads to a dependence of $\bm Q$ on the spin.

In CM, the position vector satisfying Eq. (\ref{positioncommutation})
is the radius vector $\bm R$.
For a free Dirac particle, the most straightforward way for a
determination of the position and spin operators in any representation
is the use of the FW representation as a starting point. The reason
is a deep similarity between the classical Hamiltonian (\ref{HK})
(which is spin-independent for a free particle) and the corresponding
FW Hamiltonian \cite{FW}
\begin{equation}
{\cal H}_{FW}=\beta\sqrt{m^2+\bm p^2},
\qquad \bm p\equiv-i\hbar\frac{\partial}{\partial\bm r}.
\label{HamFW}
\end{equation}
In addition, the lower spinor of the FW wave function $\Psi_{FW}$
is equal to zero if the total particle energy is positive. The
Hamiltonian (\ref{HamFW}) results from the FW transformation of
the Dirac Hamiltonian
\begin{equation}
{\cal H}_{D}=\beta m+\bm\alpha\cdot\bm p.
\label{HamD}
\end{equation}

The remaining operators read \cite{JordanMuku}
\begin{equation}
\bm j=\bm l+\bm s, \quad \bm l\equiv\bm q\times\bm p,
\quad \bm K=\frac12(\bm q{\cal H}+{\cal H}\bm q)
-\frac{\bm s\times\bm p}{\beta m+{\cal H}}-t\bm p,
\label{tomFW}
\end{equation}
where $\bm q$ is the position operator.

The operators being counterparts of fundamental classical variables
should satisfy the relations [cf. Eqs. (\ref{Poisson}) -- (\ref{spinOAMbrackets})]
\begin{equation}
\begin{array}{c}
[p_i,p_j]=0,\quad [p_i,{\cal H}]=0,\quad [j_i,{\cal H}]=0,
\quad \left[j_i,j_j\right]=ie_{ijk}j_k,\quad [j_i,p_j]=ie_{ijk}p_k,\\
\left[j_i,K_j\right]=ie_{ijk}K_k,	\quad
\left[K_i,{\cal H}\right]=ip_{i},\quad \left[K_i,K_j\right]=-ie_{ijk}j_k,
\quad [K_i,p_j]=i\delta_{ij}{\cal H},
\\
\left[q_i,K_j\right]=\frac12\left(q_j\left[q_i,{\cal H}\right]
+\left[q_i,{\cal H}\right] q_j\right)-it\delta_{ij},\quad
\left[q_i,p_j\right]=i\delta_{ij},\quad [q_i,j_j]=ie_{ijk}q_k,\quad \\
\left[q_i,s_j\right]=0,\quad \left[s_i,p_j\right]=0,\quad \left[l_i,s_j\right]=0,
\quad \left[l_i,l_j\right]=ie_{ijk}l_k,\quad \left[s_i,s_j\right]=ie_{ijk}s_k,
\end{array}
\label{commutator}
\end{equation}
\begin{equation}
\left[q_i,q_j\right]=0. \label{kcommfq}
\end{equation}

Let us first consider the set of operators
$\bm p, {\cal H}_D, \bm j, \bm K, \bm q, \bm s_D$,
where $\bm s_D=\hbar\bm\Sigma/2$ and all these operators are
defined in the Dirac representation (in particular, the position
operator is the Dirac radius vector $\bm r$). Some of commutators
in Eq. (\ref{commutator}) which contain $\bm K$ are not satisfied
by these operators. This fact follows from a noncoincidence of
the position operator in the Dirac representation with $\bm r$
which has been shown for the first time in Ref. \cite{Pryce}.

A consideration of the set of operators
$\bm p, {\cal H}_{FW}, \bm j, \bm K, \bm q, \bm s$ defined in
the FW representation leads to an opposite conclusion. In this
representation, the definition of $\bm s$ is the same
($\bm s=\hbar\bm\Sigma/2$) and the position operator $\bm q$
is equal to the FW radius vector $\bm x$. We can check that
Eqs. (\ref{commutator}) and (\ref{kcommfq}) are now satisfied.
Thus, the counterparts of the classical Hamiltonian, the position
vector, the orbital angular momentum (OAM), and the spin are the
operators ${\cal H}_{FW},\, \bm x,\, \bm x\times\bm p$, and
$\hbar\bm\Sigma/2$ \emph{defined in the FW representation}.
The operators $\bm p$ and $\bm J$ are not changed by the
transformation from the Dirac representation to the FW one
and the counterpart of the classical variable $\bm K$ is the
FW operator (\ref{tomFW}) with $\bm q=\bm x$.

Evidently, the Hamiltonian (\ref{HamFW}) commutes with the
OAM and spin operators.

The choice between the definitions of fundamental operators
in the Dirac and FW representations becomes evident when the
commutators of the Hamiltonian with the position operator are
considered. The corresponding Poisson bracket following from
Eq. (\ref{HK}) is equal to
$$
\left\{H,\bm Q\right\}=-\frac{\bm p}{H}.
$$
Since
the center-of-charge velocity is defined by
$$
\bm V\equiv\frac{d\bm Q}{dt}=\frac{\partial H}{\partial \bm p}=\frac{\bm p}{H},
$$
we obtain the relation
\begin{equation}
\bm V=\frac{\bm p}{H}.
\label{Pb2}
\end{equation}
The commutators are given by
\begin{equation}
\begin{array}{c}
\left[{\cal H}_D,\bm r\right]=-i\frac{d\bm r}{dt}\equiv -i\bm v_D=-i\bm\alpha,\\
\left[{\cal H}_{FW},\bm x\right]=-i\frac{d\bm x}{dt}\equiv -i\bm v_{FW}=-i\frac{\bm p}{{\cal H}_{FW}}.
\label{DFW2}
\end{array}
\end{equation}
Equations (\ref{Pb2}) and (\ref{DFW2}) show that only the FW
operators are the quantum-mechanical counterparts of the corresponding
classical variables. The connection between the velocity and momentum
operators is closely related to the problem of \textit{Zitterbewegung} considered
in Subsec. \ref{ZitterbewegungQM}.
An importance of the proportionality between the velocity and momentum
operators has been noted in Refs. \cite{OConnell,Mattews,Costella}.

Of course, the counterparts of the fundamental classical variables can
be determined in any representation. In the Dirac representation, they
are defined by the transformation of the corresponding FW operators
\cite{FW,JordanMuku,Foldy56,Foldy61,AcharyaSudarshan,Bacry88}. This
transformation is inverse with respect to the FW one and is performed
by the operator $U_{FW}^{-1}$. If we denote by $A$ any fundamental
operator in the FW representation, the same operator in the Dirac
representation is equal to $U_{FW}^{-1}AU_{FW}$. Thus, the counterparts
of the fundamental classical variables in the Dirac representation read
\begin{equation}
\begin{array}{c}
\bm P\rightarrow\bm p=\bm p_D=\bm p_{FW},\quad H\rightarrow {\cal H}_D
=U_{FW}^{-1}{\cal H}_{FW}U_{FW},\\
\bm J\rightarrow\bm j=\bm j_D=\bm j_{FW},\quad \bm Q\rightarrow\bm q
=\bm X=U_{FW}^{-1}\bm xU_{FW},\\
\bm L\rightarrow\bm l=\bm l_D=U_{FW}^{-1}\bm x\times\bm pU_{FW}
=\bm X\times\bm p,\quad \bm S\rightarrow\bm s=\bm {\mathcal{S}}
=\frac{\hbar}{2}U_{FW}^{-1}\bm\Sigma U_{FW},\\
\bm K\rightarrow\bm K_D= U_{FW}^{-1}\left[\frac12(\bm x{\cal H}_{FW}
+{\cal H}_{FW}\bm x)
-\frac{\bm s\times\bm p}{m+{\cal H}_{FW}}-t\bm p\right]U_{FW}\\
=\frac12(\bm X{\cal H}_{D}+{\cal H}_{D}\bm X)
-\frac{\bm{\mathcal{S}}\times\bm p}{m+{\cal H}_{D}}-t\bm p.
\label{Diracfundamentals}
\end{array}
\end{equation}
Here the operators of the position (``mean position'' \cite{FW}) and
the spin (``mean spin angular momentum'' \cite{FW}) in the Dirac
representation are equal to \cite{Pryce,FW,dVFor}
\begin{equation}
\bm q=\bm X=\bm r -\frac{\bm\Sigma\times\bm p}{2\epsilon(\epsilon+m)}
+\frac{i\bm\gamma}{2\epsilon}-\frac{i(\bm\gamma\cdot\bm p)\bm p}{2\epsilon^2(\epsilon+m)},
\label{meanpos}
\end{equation}
\begin{equation}
\bm{\mathcal{S}}=\frac{m}{2\epsilon}\bm\Sigma-i\frac{\bm\gamma\times\bm p}{2\epsilon}
+\frac{\bm p(\bm \Sigma\cdot\bm p)}{2\epsilon(\epsilon+m)},
\quad \epsilon=\sqrt{m^2+\bm p^2}.
\label{menspin}
\end{equation}
We underline that the conventional spin operator corresponding to
the classical rest-frame spin commutes with the OAM operator, the
Hamiltonian, and the position and momentum operators \emph{in any
representation}. For any operators satisfying the relation $C_{FW}=[A_{FW},B_{FW}]$,
$$
\begin{array}{c}
C_{D}=U_{FW}^{-1}C_{FW}U_{FW}=U_{FW}^{-1}(A_{FW}B_{FW}-B_{FW}A_{FW})U_{FW}
= U_{FW}^{-1}A_{FW}U_{FW}U_{FW}^{-1} B_{FW}U_{FW}\\
-U_{FW}^{-1}B_{FW}U_{FW}U_{FW}^{-1}A_{FW}U_{FW}
=[A_{D},B_{D}].
\end{array}
$$

The validity of the above-mentioned results on the position, spin,
and other fundamental operators in the Dirac and FW representations
has been demonstrated by \emph{numerous} methods. Newton and Wigner
\cite{NewtonWigner} (see also Ref. \cite{Wightman}) have investigated
localized
states for elementary systems. They have shown that the operator
(\ref{meanpos}) is the only position operator (with commuting components)
in the Dirac theory which has localized eigenfunctions in the manifold
of wave functions describing positive-energy states \cite{NewtonWigner}.
Therefore, the operator (\ref{meanpos}) is called the Newton-Wigner
(NW) position operator.

It is important that the deep similarity between the fundamental
classical variables and the corresponding FW operators
does not disappear for different definitions of the position operator.
It has been shown still in Ref. \cite{Pryce} that definitions of this
operator violating the relation $[q_i,q_j]=0$ are possible for spinning
particles. The subsequent investigations
\cite{SuttorpDeGroot,DeKerfBauerle,KhPom,PomKh,Rivas,Stepanov,Costaetal}
have confirmed the possibility of position operators with noncommutative
components for spinning particles. However, the position operator with
commutative components should satisfy Eqs. (\ref{tomFW})-(\ref{meanpos}).

The fundamental conclusion that the NW position operator $\bm q$
and the radius vector in the FW representation $\bm x$ are identical
has been confirmed in many papers
\cite{FGursey,Fronsdal,Bacry,OConnell,Kalnayetal,Foldy,BGS,Mattews,Costella}.
Some of them have been fulfilled by different methods.
In particular, the extended-type position operator has been proposed
in Ref. \cite{Kalnayetal}
and definite relations between the velocity ($\dot{\bm q}$) and momentum
operators have been introduced in Refs. \cite{OConnell,Mattews}.

The equivalence of the classical spin $\bm S$ and the FW mean-spin
operator has also been shown in Refs.
\cite{FG,FGursey,BGS,Mattews,Costella,Ryder,Caban,Bauke,CKT}. A rather
important result has been obtained by Fradkin and Good \cite{FG}.
They not only have confirmed Eq. (\ref{menspin}) for the spin operator
in the Dirac representation but also have demonstrated that the result
obtained by Foldy and Wouthuysen remains valid for a Dirac particle in
electric and magnetic fields.
The FW mean-spin operator defines the rest-frame spin \cite{FG}, while
the use of the four-component spin operator $a^\mu$ is also admissible \cite{FG,BLP}.
This operator is orthogonal to the four-momentum
one ($a^\mu p_\mu=0$) and is defined by
\begin{equation}
a^{\mu}=\frac{1}{2m}e^{\alpha\beta\nu\mu}p_\alpha S_{\beta\nu},
\label{fourspintens}
\end{equation}
where $S_{\beta\nu}$ is the antisymmetric spin tensor.
Evidently, the four-component spin operator
\begin{equation}
a^\mu=(a^0,\bm a)=\left(\frac{\bm p\cdot\bm s}{m},\,
\bm s+\frac{\bm p(\bm s\cdot\bm p)}{m(\epsilon+m)}\right).
\label{fourspin}
\end{equation}
also commutes with the Hamiltonian. However, $\bm a$ cannot be the
conventional spin operator because it does not commute with the
operators $\bm q,\bm l$ and does not satisfy other commutative
relations [see Eq. (\ref{commutator})]. Certainly, the rest-frame
spin $\bm s$ is invariant relative to Lorentz boosts.

We also note important analyzes presented in Refs.
\cite{Caban,Bauke,HehlArXiv}. It has been shown in Refs.
\cite{Caban,Bauke} that the operator defining the conventional
spin in the Dirac representation is the mean-spin operator
(\ref{menspin}) introduced by Foldy and Wouthuysen. It has been
concluded in Ref. \cite{HehlArXiv} that the Gordon decomposition
of the energy momentum and spin currents of the Dirac electron
corresponds to the FW transformation of its wave function.

Dirac particles in (1+1) dimensions have been considered in
Refs. \cite{Toyamaetal,AlonsoDeVincenzo}. In the FW representation,
wave packets described by the (1+1)-dimensional Dirac equation
also behave much more like a classical particle than in the Dirac
representation \cite{Toyamaetal,AlonsoDeVincenzo}.

Thus, the correct forms of conventional operators of the position
and spin of a free Dirac particle are defined by Eqs. (\ref{meanpos})
and (\ref{menspin}) in the Dirac representation. These operators are
equal to the radius vector $\bm x$ and to the spin operator
$\hbar\bm\Sigma/2$ in the FW representation.

\section{Classical limit for a Dirac fermion and spin-0 and
spin-1 bosons in external fields}\label{Comparison}

In the precedent section, we have analyzed free particles and
this analysis is fully based on
the results obtained many years ago. However, the contemporary
development of theory of the FW transformation
allows us to put forward important arguments in favor of the similarity between
the classical position and spin and the corresponding operators
in the FW representation. This section, unlike the precedent one,
is devoted to a consideration of particles \emph{in external fields}.

Relativistic methods giving compact relativistic FW Hamiltonians
for any energy allow one to establish a
direct connection between classical and quantum-mechanical
Hamiltonians. To find this connection, it is
convenient to pass to the classical limit of relativistic
quantum-mechanical equations. Importantly, this procedure
is very simple in the FW representation. When the conditions
of the Wentzel-Kramers-Brillouin approximation
are satisfied, the classical limit can be obtained by replacing
the FW operators with the respective classical variables
\cite{JINRLett12}. This property leads to the conclusion that
the quantum-mechanical counterparts of the classical variables
are the corresponding operators \emph{in the FW representation}.

Let us begin the analysis of Dirac particle interactions with
external fields from the result obtained in Ref. \cite{FG}.
In this paper, the equation of spin motion has been derived
\emph{in the Dirac representation} and its classical limit
has been obtained. A particle with an AMM has been considered
and the initial Dirac-Pauli equation [Eq. (\ref{eqDPgen}) with $d=0$]
has been used. In the classical limit, Fradkin and Good have
obtained the equation \cite{FG} coinciding with the famous
classical Thomas-Bargmann-Michel-Telegdi (T-BMT) one \cite{Thomas,BMT}.
The presence of the Thomas term shows that the both equations
are derived for the rest-frame spin $\bm{\mathcal{S}}$ but not
for the spin in the laboratory frame or in the instantaneously accompanying
one. The distinction between the rest frame and the instantaneously
accompanying one can be made only for an accelerated particle.

The use of the FW representation leads to the same conclusion.
The relativistic FW Hamiltonian for the Dirac particle with the
AMM and EDM obtained in Ref. \cite{RPJ} is given by Eqs.
(\ref{FWHamEDM})-(\ref{EDMeq12}). To compare the position
and spin operators with their classical counterparts, we can use
the weak-field approximation and can disregard terms proportional
to $\hbar^2$ and describing contact interactions. When the fields
are uniform, the gauge $\Phi=-\bm E\cdot\bm x,\,\bm A=(\bm B\times\bm x)/2$
can be used.
In this case, the Hamiltonian (\ref{FWHamEDM}) takes the form
\begin{equation}
\begin{array}{c}
{\cal H}_{FW}= \beta\sqrt{m^2+\left(\bm p-\frac{e}{2}\bm B\times\bm x\right)^2}
-e\bm E\cdot\bm x+\bm\Omega\cdot\bm s,\quad
\bm\Omega=\bm{\Omega}_{MDM}+\bm{\Omega}_{EDM},\\
\bm{\Omega}_{MDM}=\frac{e}{m}\left[-\beta\left(\frac{m}{\epsilon}
+a\right){\bm B}+\beta\frac{a}{\epsilon(\epsilon+m)}({\bm p}\cdot{\bm B}){\bm p}
+\frac{1}{\epsilon}\left(\frac{m}{\epsilon+m}+a\right){\bm p}\times{\bm E}\right],\\
\bm{\Omega}_{EDM}=-\frac{e\eta}{2m}\left[\beta\bm
E-\beta\frac{({\bm p}\cdot\bm E){\bm p}}{\epsilon(\epsilon+m)}+\frac{{\bm p} \times\bm
B}{\epsilon}\right],\quad \bm s=\frac{\bm\Sigma}{2},\quad \epsilon=\sqrt{m^2+\bm p^2},
\end{array}
\label{eqFWEDM12}
\end{equation}
where $a=(g-2)/2$, $g=4mc(\mu_0+\mu')/(e\hbar)$, and $\eta=4mcd/(e\hbar)$
is the ``gyroelectric'' factor corresponding to $g$. The matrix $\beta$
may be removed if one considers positive-energy states and disregards
the zero lower spinor. The equation of spin motion is given by
\begin{equation}
2\frac{d\bm s}{dt}=\frac{d\bm\Sigma}{dt}=\bm\Omega\times\bm\Sigma.
\label{eqFWspinn}
\end{equation}
The operator of the angular velocity of spin rotation $\bm\Omega$ has
the two parts, $\bm{\Omega}_{MDM}$ and $\bm{\Omega}_{EDM}$, defining
the contributions of the magnetic dipole moment and the EDM, respectively.

The related relativistic FW Hamiltonians derived in
Refs. \cite{electrodynamics,ChiouChen,JMP,PRA2008} agree with the
Hamiltonian (\ref{eqFWEDM12}). We underline that the method of the
relativistic FW transformation used in Ref. \cite{ChiouChen}
substantially differs from that applied in other above-mentioned works.
The operator $\bm{\Omega}_{MDM}$ is in compliance with the operator
of the angular velocity of spin motion in the Dirac representation
obtained in Ref. \cite{FG}.

We can now compare the Hamiltonian (\ref{eqFWEDM12}) and the equation
of spin motion (\ref{eqFWspinn}) with their classical counterparts.
In the same approximation, the classical Hamiltonian of a spinning
particle in uniform electric and magnetic fields has the form
\begin{equation}
\begin{array}{c}
H= \sqrt{m^2+\left(\bm P-\frac{e}{2}\bm B\times\bm R\right)^2}
-e\bm E\cdot\bm R+\bm\Omega\cdot\bm S,
\end{array}
\label{eqclHam}
\end{equation}
where the angular velocity of spin rotation
$\bm\Omega=\bm{\Omega}_{MDM}+\bm{\Omega}_{EDM}$ is defined
by (see Refs. \cite{Nelsonm,FukuyamaSilenko,PhysScr} and
references therein)
\begin{equation}
\begin{array}{c}
\bm{\Omega}_{MDM}=\frac{e}{m}\left[-\left(\frac{m}{\varepsilon}
+a\right){\bm B}+\frac{a}{\varepsilon(\varepsilon+m)}({\bm P}\cdot{\bm B}){\bm P}
+\frac{1}{\varepsilon}\left(\frac{m}{\varepsilon+m}+a\right){\bm P}\times{\bm E}\right],\\
\bm{\Omega}_{EDM}=-\frac{e\eta}{2m}\left[\bm
E-\frac{({\bm P}\cdot\bm E){\bm P}}{\varepsilon(\varepsilon+m)}+\frac{{\bm P}\times\bm
B}{\varepsilon}\right],\quad \varepsilon=\sqrt{m^2+\bm P^2}.
\end{array}
\label{eqclOmg}
\end{equation}

The comparison of Eqs. (\ref{eqFWEDM12}) and (\ref{eqFWspinn})
with (\ref{eqclHam}) and (\ref{eqclOmg}) unambiguously shows that
the classical counterparts of the FW position operator $\bm x$
and the FW spin operator $\bm s=\hbar\bm\Sigma/2$ are the radius
vector $\bm R$ and the rest-frame spin $\bm S$, respectively.
This is a strong argument in favor of the statements that
the position operators are the FW radius vector $\bm x$ and
the Dirac operator (\ref{meanpos}) and that the conventional
spin operators are the FW operator $\hbar\bm\Sigma/2$ and the
Dirac operator (\ref{menspin}). In this section, Eqs. (\ref{meanpos})
and (\ref{menspin}) define the Dirac position
and spin operators only approximately because the FW transformation
operator depends on external fields.

One can confirm these statements for a Dirac particle in
gravitational fields and noninertial frames. It has been
definitely shown in many papers devoted to this problem
\cite{PRD,PRD2,Warszawa,OST,OSTRONG,ostgrav,ostor,ostor2,OSTalkPRD,OSTotal}
that the relativistic quantum-mechanical Hamiltonians and
equations of motion in the FW representation are similar
to the corresponding classical ones.
As an example, let us consider the Dirac particle in the
general noninertial frame. This frame is characterized by
the acceleration $\bm{\mathfrak{a}}$ and the rotation with
the angular velocity $\bm\omega$. The relativistic FW
Hamiltonian reads \cite{OSTRONG}
\begin{equation}
\begin{array}{c}
{\cal H}_{FW}=\frac\beta2\left[\left(1+\frac{\bm{\mathfrak{a}}\cdot\bm
x}{c^2}\right)
\epsilon+\epsilon\left(1+\frac{\bm{\mathfrak{a}}\cdot\bm
x}{c^2}\right)\right]-\bm\omega\cdot\bm l+\frac\hbar2\bm\Omega\cdot\bm\Sigma,\\
\bm\Omega= \beta\frac{\bm{\mathfrak{a}}\times\bm
p}{\epsilon+mc^2}-\bm\omega,\qquad\epsilon
=\sqrt{m^2c^4+c^2\bm p^2},\qquad\bm l=\bm x\times\bm p.
\end{array}
\label{Hamgnif}
\end{equation}
Let us stress that Eq.
(\ref{Hamgnif}) has been derived for the strong kinematical effects when
the ratios $|\bm{\mathfrak{a}}\cdot \bm x|/c^2$ and $|\bm\omega\times\bm x|/c$
are not small.

The corresponding classical Hamiltonian can be obtained with a
substitution of the metric of the general noninertial frame
into Eq. (3.18) from Ref. \cite{OSTRONG}:
\begin{equation}
\begin{array}{c}
H=\left(1+\frac{\bm{\mathfrak{a}}\cdot\bm
R}{c^2}\right)\varepsilon-\bm\omega\cdot\bm L+\bm\Omega\cdot\bm S,\\
\bm\Omega=\frac{\bm{\mathfrak{a}}\times\bm
P}{\varepsilon+mc^2}-\bm\omega,\qquad\varepsilon=\sqrt{m^2c^4+c^2\bm P^2}.
\end{array}
\label{Hamltni}
\end{equation}
It follows from Eqs. (\ref{Hamgnif}) and (\ref{Hamltni}) that the
position and spin operators are the FW operators $\bm x$ and
$\bm s=\hbar\bm\Sigma/2$ and the Dirac operators (\ref{meanpos})
and (\ref{menspin}), respectively.

Because of the unification of relativistic QM in the FW
representation \cite{PRDProca}, similar statements can be made for
spin-0 and spin-1 particles. In connection with this unification,
we can mention
the existence of bosonic symmetries of the standard Dirac equation
\cite{Simulik,Simulik1,Simulik2,Simulik3,Simulik4,Simulik5,Simuliknew}.
When terms proportional to $\hbar^2$ are disregarded and the weak-field
approximation is used, the relativistic Hamiltonian for a spin-0 particle
in the uniform electric and magnetic fields has the form \cite{TMP2008}
\begin{equation}
\begin{array}{c}
{\cal H}_{FW}= \rho_3\sqrt{m^2+\left(\bm p-\frac{e}{2}\bm B\times\bm x\right)^2}
-e\bm E\cdot\bm x, \end{array}
\label{eqFWsca}
\end{equation}
where $\rho_3$ is the corresponding Pauli matrix acting
on a two-component wave function. On the same conditions,
the relativistic Hamiltonian for a spin-1 particle with
the AMM and EDM in the uniform electric and magnetic fields
is given by \cite{PhysRevDspinunit}
\begin{equation}
\begin{array}{c}
{\cal H}_{FW}= \beta\sqrt{m^2+\left(\bm p-\frac{e}{2}\bm B\times\bm x\right)^2}
-e\bm E\cdot\bm x+\bm\Omega\cdot\bm s^{(1)},\quad \bm\Omega=\bm{\Omega}_{MDM}+\bm{\Omega}_{EDM},\\
\bm{\Omega}_{MDM}=\frac{e}{m}\left[-\beta\left(\frac{m}{\epsilon}+a\right){\bm B}
+\beta\frac{a}{\epsilon(\epsilon+m)}({\bm p}\cdot{\bm B}){\bm p}
+\frac{1}{\epsilon}\left(\frac{m}{\epsilon+m}+a\right){\bm p}\times{\bm E}\right],\\
\bm{\Omega}_{EDM}=-\frac{e\eta}{2m}\left[\beta\bm
E-\beta\frac{({\bm p}\cdot\bm E){\bm p}}{\epsilon(\epsilon+m)}+\frac{{\bm p} \times\bm
B}{\epsilon}\right],\quad \epsilon=\sqrt{m^2+\bm p^2},
\end{array}
\label{eqFWEDMunit}
\end{equation}
where
$$
\beta=\left(\begin{array}{cc} \mathfrak{I} & 0 \\
0 & -\mathfrak{I} \end{array}\right),
$$
\begin{equation}
\bm{s}^{(1)}=\left(\begin{array}{cc} \bm{S}^{(1)} & 0 \\
0 & \bm{S}^{(1)} \end{array}\right),
\label{eqspinunit}
\end{equation}
$a=(g-2)/2$, $g=2mc\mu/(e\hbar)$, and $\eta=2mcd/(e\hbar)$.
Here $\bm{S}^{(1)}=(S^{(1)}_i)$ is the conventional $3\times3$
spin matrix for spin-1 particles and $\mathfrak{I}$ is the
$3\times3$ unit matrix. The wave function has six components.
The matrix $\beta$ may be removed if one considers positive-energy
states and disregards the zero lower spinor-like part of the
FW wave function.

Evidently, the definition of the position operator as the radius
vector in the FW representation remains valid for spin-0 and
spin-1 particles. The use of the FW transformation for a description
of a relativistic spin-0 particle in gravitational fields and
noninertial frames \cite{Honnefscalar} confirms this definition
of the position operator. The fundamental spin operator for a
spin-1 particle in the FW representation is the matrix (\ref{eqspinunit}).

The basic role of the FW representation in nonstationary QM has
been shown 
in Ref. \cite{ExpectationValue}.
The classical time-dependent energy corresponds to the time-dependent
expectation value of the energy operator. The latter is the Hamiltonian
in the Schr\"{o}dinger QM and the FW representation (but not in the
Dirac representation) \cite{ExpectationValue}. The energy expectation
values are defined by \cite{ExpectationValue}
\begin{equation}
E(t)=\int{\Psi_{FW}^\dag(\bm r,t){\cal H}_{FW}(t)\Psi_{FW}(\bm r,t)dV}.
\label{EEVFW}
\end{equation}
In the Dirac representation,
\begin{equation}
E(t)=\int{\Psi_{D}^\dag(\bm r,t)\widetilde{{\cal H}}(t)\Psi_{D}(\bm r,t)dV},
\label{EEVD}
\end{equation}
where $\widetilde{{\cal H}}(t)$ is the energy operator which defines
the energy expectation values by averaging. It does not coincide with
the Dirac Hamiltonian and is equal to \cite{ExpectationValue}
\begin{equation}
\begin{array}{c}
\widetilde{{\cal H}}(t)={\cal H}_D+\frac{e\hbar}{8}\left\{\frac{1}{\epsilon'(\epsilon'
+m)},\left[-i\{\epsilon',\bm{\gamma}\cdot\dot{\bm A}\}
-2im\bm{\gamma}\cdot\dot{\bm A}+\bm\Sigma\cdot(\bm\pi
\times\dot{\bm A}-\dot{\bm A}\times\bm\pi)\right]\right\}\\
+i\frac{e\hbar}{8}\left\{\frac{1}{{\epsilon'}^2(\epsilon'
+m)},\left[(\bm{\pi}\cdot\dot{\bm A})(\bm{\gamma}\cdot{\bm\pi})
+(\bm{\gamma}\cdot{\bm\pi})(\dot{\bm A}\cdot\bm{\pi})\right]\right\},
\end{array}
\label{TrHamfnEEV}
\end{equation}
where ${\cal H}_D$ is the Dirac Hamiltonian, $\epsilon'=\sqrt{m^2+\bm\pi^2}$,
and dots denote time derivatives.

The contribution to the energy expectation values given by the two last terms
in Eq. (\ref{TrHamfnEEV}) can be rather important.
This equation shows that the Dirac Hamiltonian does not correspond to the
classical one in the nonstationary case \cite{ExpectationValue}.

In fact, the difference between the position operator (\ref{meanpos})
and the radius vector $\bm r$ in the Dirac representation is very important.
The assumption that $\bm r$ is the true Dirac position operator leads to the
misleading conclusion that the quantity $\varrho_D=\Psi_D^{\dag}(\bm r)\Psi_D(\bm r)$
is the probability density that the particle is at the point $\bm r$ and
the quantity $e\Psi_D^{\dag}(\bm r)\Psi_D(\bm r)$ describes the electron
charge distribution. This assumption also results in a calculation of
incorrect expectation values of operators. We will discuss these problems
in Sec. \ref{Discussions}.

Thus, the consideration of a Dirac particle in external fields
leads to results fully supporting the conclusions made in the
precedent section. An analysis of spin-0 and spin-1 particles
in external fields also presents arguments in favor of these conclusions.
In contrast to the results for a free particle presented in Sec.
\ref{Classical}, the particle spin motion in electric and magnetic field
is sensitive to the Thomas effect \cite{Thomas} and unambiguously shows that the
fundamental spin operator is defined in the particle rest frame.
The analysis presented excludes the possibility of a definition
of this operator in the instantaneously accompanying frame.

\section{Relativistic operators of the position and spin}\label{Operators}

The rest-frame spin $\bm s$ and the four-component one $a^\mu$ do
not exhaust the list of relativistic spin operators. The spin can
also be represented by the antisymmetric tensor (see Ref. \cite{BLP}, Sec. 29)
\begin{equation}
S^{\mu\nu}=\frac1me^{\mu\nu\alpha\beta}a_\alpha p_\beta.
\label{spintas}
\end{equation}
Similarly to the OAM, the spatial part (components $S^{ij}$) of
this antisymmetric tensor forms the three-component vector $\bm\zeta$
with the following transformation properties (see Ref. \cite{BLP}, Sec. 29):
\begin{equation}
\bm\zeta^{(0)}=\bm s, \quad \zeta_\parallel=\zeta^{(0)},
\quad \zeta_\perp=\frac\epsilon m\zeta^{(0)}, \quad \bm\zeta
=\frac\epsilon m\bm\zeta^{(0)}-\frac{(\bm\zeta^{(0)}\cdot\bm p)\bm p}{m(\epsilon+m)},
\label{spinzeta}
\end{equation}
where $\bm\zeta^{(0)}$ characterizes the particle rest frame.
Evidently, the vectors $\bm a$ and $\bm\zeta$ differ. The
quantity $\bm\zeta$ defines the three-component \emph{laboratory-frame}
spin and can be written in the form
\begin{equation}
\bm\zeta=\bm s-\frac{\bm p\times(\bm p\times\bm s)}{m(\epsilon+m)}.
\label{zetatwo}
\end{equation}

The quantities $\bm l$ and $\bm s$ forming the total angular
momentum $\bm j$ have different physical meanings. The OAM $\bm l$
is the spatial part of the antisymmetric tensor
$L^{\mu\nu}=(-\bm\kappa,-\bm l)$ with $\bm\kappa=(\bm q{\cal H}+{\cal H}\bm q)/2
-t\bm p$ and is noninvariant relative to Lorentz transformations.
The rest-frame spin $\bm s$ is invariant relative to such transformations.
It is natural to constitute the total angular momentum from spatial
parts of the two antisymmetric tensors, $L^{\mu\nu}$ and $S^{\mu\nu}$:
\begin{equation}
J^{\mu\nu}=L^{\mu\nu}+S^{\mu\nu}=x^{\mu}p^{\nu}-x^{\nu}p^{\mu}+S^{\mu\nu}.
\label{JLS}
\end{equation}
Since the spatial part of $S^{\mu\nu}$ is presented
by the vector $\bm\zeta$, the definition of this vector is analogous
to the definition of the total angular momentum $\bm j$.
Equation (\ref{tomFW}) shows that the corresponding operators
of the position and OAM should be redefined in order to avoid a
change of the operator $\bm j$:
\begin{equation}
\bm{j}=\bm l+\bm s=\bm{\mathcal{L}}+\bm\zeta,
\quad \bm{\mathcal{L}}=\bm{\mathcal{X}}\times\bm p.
\label{podeGSp}
\end{equation}
Equations (\ref{zetatwo}) and (\ref{podeGSp}) specify the
position operator $\bm{\mathcal{X}}$ \cite{Pryce,SuttorpDeGroot,DeKerfBauerle}.
In the FW representation \cite{Pryce,SuttorpDeGroot,DeKerfBauerle},
\begin{equation}
\bm{\mathcal{X}}_{FW}=\bm x +\frac{\bm s\times\bm p}{m(\epsilon+m)},
\quad \bm{\mathcal{L}}_{FW}=\bm{\mathcal{X}}_{FW}\times\bm p,
\label{newposo}
\end{equation}
where $\bm x$ is the FW center-of-charge position operator and
the spin operator $\bm\zeta_{FW}=\bm\zeta$ is given by Eq. (\ref{zetatwo}).

In the Dirac representation
\cite{Pryce,SuttorpDeGroot},
\begin{equation}
\bm{\mathcal{X}}_D=\bm r +i\left[\frac{\bm\gamma}{2m}
-\frac{(\bm\gamma\cdot\bm p)\bm p}{2m\epsilon^2}\right],
\quad \bm{\mathcal{L}}_D=\bm{\mathcal{X}}_D\times\bm p,
\label{XDiracr}
\end{equation}
\begin{equation}
\bm\zeta_D=\frac{\bm\Sigma}{2}-i\frac{\bm\gamma\times\bm p}{2m},
\label{zDiracr}
\end{equation} where $\bm r$ is the Dirac position operator.
Certainly,
$\bm{\mathcal{X}}_D\times\bm p+\bm\zeta_D=\bm{\mathcal{X}}_{FW}\times\bm p+\bm\zeta=\bm j$.
The choice of the laboratory-frame operators $\bm{\mathcal{X}}$ and $\bm{\zeta}$ have been characterized as a preferable choice of the position and spin operators in Refs. \cite{Deriglazov1,Deriglazov2}. These operators correspond to the case (d) in Pryce's classification \cite{Pryce}.

The Dirac and FW position operators have the same form,
$\bm r=(x,y,z)$ and $\bm x=(x,y,z)$. However, they define
different physical quantities, see Eq. (\ref{meanpos}).
Only the FW position operator (``mean position operator'' \cite{FW})
is the quantum-mechanical counterpart of the classical position
variable $\bm R=(X,Y,Z)$ \cite{FW}.

In the framework of \emph{covariant} spin physics, the operator
$\bm q$ should define the position of the center
of mass. In this case, its determination is based on the use of
the \emph{laboratory-frame} spin $\bm\zeta$ and OAM $\bm{\mathcal{X}}\times\bm p$
and of the corresponding position operator $\bm{\mathcal{X}}$.
Therefore, just the operator $\bm q=\bm{\mathcal{X}}$ characterizes
the center of mass of a particle. By virtue of Eq. (\ref{newposo}),
the positions of the center of mass and the center of charge
(or the mass point for an uncharged particle, see Sec. \ref{Classical}) differ.

The corresponding classical variables are very similar.
In classical physics, the center-of-mass position, the
corresponding OAM, and the laboratory-frame spin are given by
\begin{equation}
\bm{\mathcal{X}}=\bm R +\frac{\bm S\times\bm P}{m(H+m)},
\quad \bm{\mathcal{L}}=\bm{\mathcal{X}}\times\bm P,\quad \bm\zeta
=\bm S-\frac{\bm P\times(\bm P\times\bm S)}{m(H+m)}.
\label{COMOAMLFS}
\end{equation}

More recent investigations \cite{KhPom,PomKh,Obzor,Bauke,CKT} have confirmed
the validity of results obtained in Refs. \cite{Pryce,SuttorpDeGroot,DeKerfBauerle}
and have given important substantiations of the meaning of the
operator $\bm{\mathcal{X}}$. In particular, the shift of the center
of mass relative to the center of charge manifests itself in
spin-orbit and spin-spin effects in gravitational interactions \cite{KhPom}.
The shift $\bm{\mathcal{X}}_{FW}-\bm x=\bm s\times\bm p/[m(\epsilon+m)]$
naturally appears in the original quantum-mechanical approach expounded in Refs. \cite{PomKh,Obzor}.

Since Eq. (\ref{JLS}) is covariant and it leads \emph{only} to the
relation $\bm{j}=\bm{\mathcal{L}}+\bm\zeta$ (but not to any different relation
like $\bm{j}=\bm l+\bm s$), the covariant spin physics should be based
on the operators $\bm{\mathcal{X}}, \bm{\mathcal{L}},\bm\zeta$. In the
general case, equations of momentum and spin dynamics obtained with the
operators  and \emph{any different} set of fundamental operators
(e.g., $\bm x, \bm l, \bm s$) are noncovariant. This fact (first noted in
Refs. \cite{KhPom,PomKh,Obzor}) does not mean that the use of different
sets can result in some mistakes. In particular, the correct utilization
of the conventional operators $\bm x, \bm l, \bm s$ is ensured by
Eq. (\ref{podeGSp}). While equations of motion obtained with these
operators can be noncovariant, such a noncovariance does not lead to any fallacy.

The Pauli-Lubanski four-vector (see Refs. \cite{Caban,CKT})
\begin{equation}
W^{\mu}=\frac12e^{\alpha\beta\nu\mu}p_\alpha J_{\beta\nu}
\label{PauliLubanski}
\end{equation}
is also widely used as a relativistic spin operator.
It is easy to check that the four-vectors $ma^\mu$ and $W^\mu$ are equivalent.
The tensor of the total angular momentum is given by Eq. (\ref{JLS}).
The tensor of the OAM does not contribute to the Pauli-Lubanski vector.
For an extended object like an atom, the spin tensor involves the \emph{internal}
OAM (for example, the OAM of an electron in an atom). In this case,
Eqs. (\ref{fourspintens}), (\ref{JLS}), and (\ref{PauliLubanski})
lead to the relation
\begin{equation}
W^{\mu}=ma^{\mu}.
\label{PauliLubanskispin}
\end{equation}

In the FW representation,
\begin{equation}
W^\mu=(W^0,\bm W)=\left(\frac{\bm p\cdot\bm\Sigma}{2},\,
\frac{m\bm\Sigma}{2}+\frac{\bm p(\bm p\cdot\bm\Sigma)}{2(\epsilon+m)}\right).
\label{PLsFW}
\end{equation}

A spinning particle is characterized by the two Casimir
invariants (Casimir operators of the Poincar\'{e}
group):
\begin{equation}
p^{\mu}p_\mu=m^2,\qquad
W^{\mu}W_{\mu}=-m^2\bm s^2=-m^2s(s+1)\mathcal{I},
\label{Casimirinv}
\end{equation} where $\mathcal{I}$ is the unit matrix.
It has been noted at the end of Sec. \ref{Classical} that the rest-frame
spin $\bm s$ is invariant relative to Lorentz boosts. Therefore,
the square of the spin operator is the Casimir
invariant and the Lorentz scalar and $\bm s$ is the correct spin operator.

It has been obtained in Ref. \cite{Caban} that the only spin
operator satisfying the required commutation relations has the form
\begin{equation}
\bm s'=\frac1m\left(\frac{|p_0|}{p_0}\bm W-W_0\frac{\bm p}{|p_0|+m}\right)
=\frac1m\left(\bm W-W_0\frac{\bm p}{p_0+m}\right).
\label{spinfin}
\end{equation}
The total energy is expected to be positive. It
has been noted in Ref. \cite{Caban} that this operator is \emph{equivalent}
to the rest-frame spin operator $\bm s$. In Ref. \cite{Caban},
nevertheless, the operators $\bm s$ and $\bm s'$ are defined by
different formulas. The use of Eqs. (\ref{fourspin}),
(\ref{PauliLubanskispin}), and (\ref{spinfin}) shows that
$\bm s'=\bm s$. This result has been first obtained by
Ryder \cite{Ryder} (see also Ref. \cite{CKT}). Therefore, the transformation of the
operator $\bm s'$ to the Dirac representation leads to the
operator (\ref{menspin}).
The operator (\ref{spinfin}) is also useful in the
quantum field theory \cite{Bogoliubov}.

In Refs. \cite{Thaller,BliokhDN,Czachor,Bliokh2017,CKT}, the
projected spin operator has been considered. It is possible
to project some operators onto positive- and negative-energy
subspaces, eliminating the cross terms
corresponding to the electron-positron transitions. In
particular, the projected radius vector operator is given by
\cite{Thaller,Czachor,Bliokh2017,CKT}
\begin{equation}
\bm{\mathfrak{R}}=\Pi^+\bm r\Pi^++\Pi^-\bm r\Pi^-,
\label{projR}
\end{equation}
where the projectors are given by
$$\Pi^\pm=\frac12U_{FW}^\dag(1\pm\beta)U_{FW}
=\frac12\left(1\pm\beta\frac{m}{\epsilon}\right)\pm\frac{\bm\alpha\cdot\bm p}{2\epsilon}.$$

In the Dirac and FW representations, the projected operators of the radius
vector (position) and spin are equal to
\begin{equation}
\bm{\mathfrak{R}}_D=\bm r-\frac{\bm\Sigma\times\bm p}{2\epsilon^2}
+i\frac{m\bm\gamma}{2\epsilon^2}, \qquad
\bm{\mathfrak{R}}_{FW}=\bm x-\frac{\bm\Sigma\times\bm p}{2\epsilon(\epsilon+m)}
\label{poDi}
\end{equation}
and
\begin{equation}
\bm{\mathfrak{S}}_D=\frac{1}{2\epsilon^2}\left[m^2\bm\Sigma
+\bm p(\bm p\cdot\bm\Sigma)-im\bm\gamma\times\bm p\right],\qquad
\bm{\mathfrak{S}}_{FW}=\frac{m\bm\Sigma}{2\epsilon}
+\frac{\bm p(\bm p\cdot\bm\Sigma)}{2\epsilon(\epsilon+m)},
\label{poDis}
\end{equation}
respectively \cite{Thaller,BliokhDN,Czachor,Bliokh2017,CKT}. The projected
OAM operator is given by \begin{equation}
\bm{\mathfrak{L}}_D=\bm{\mathfrak{R}}_{D}\times\bm p, \qquad
\bm{\mathfrak{L}}_{FW}=\bm{\mathfrak{R}}_{FW}\times\bm p.
\label{poOAM}
\end{equation}

Despite the assertion in Ref. \cite{Bliokh2017} that the projected
spin operator ``corresponds to the spatial part of the
Pauli-Lubanski four-vector'', the two vectors substantially differ
($\bm{\mathfrak{S}}_{FW}=\bm W/\epsilon\neq\bm W/m$). This conclusion
follows also from Eq. (16) in Ref. \cite{Bliokh2017}. In Pryce's
classification \cite{Pryce}, the projected operators of the position
and spin correspond to the case (c).
When our denotations are used, the classical counterpart of the
projected position operator obtained by Pryce \cite{Pryce} reads
[cf. Eqs. (\ref{tomFW}) and (\ref{poDi})]
\begin{equation}
\mathfrak{R}^i=\frac{tP^i+J^{i0}}{H},\qquad \bm{\mathfrak{R}}
=\frac{t\bm P+\bm K}{H}=\bm R-\frac{\bm S\times\bm P}{H(H+m)}.
\label{clprojR}
\end{equation}
It has been asserted in Ref. \cite{Bliokh2017} (p. 5) that the
expectation value of the projected position operator ``for a
single-electron state corresponds to the center of the probability
density (center of charge)''. However, Eqs. (\ref{poDi}) and (\ref{clprojR})
unambiguously show that the projected position depends on the spin.
Therefore, this assertion is not correct. It can be added that the
operator $\bm{\mathfrak{R}}$ substantially differs from the center-of-mass
position operator $\bm{\mathcal{X}}$ (the quantities
$\bm{\mathcal{X}}_{FW}-\bm x$ and $\bm{\mathfrak{R}}_{FW}-\bm x$ have
even opposite signs). The projected and laboratory-frame spin operators
are also substantially different. In the general case, this circumstance
results in the noncovariance of equations of motion based on the projected operators.

Nevertheless, the projected operators are needed for the description
of Berry phase effects. In this case, the noncommutativity of components
of the projected position operator is important and defines the Berry
curvature \cite{BliokhDN,Bliokh2017,XiaoDuval,RevModPhys,Gosselin,BliokhBerry}.

Thus, we can select three sets of fundamental FW operators,
$\bm x, \bm l, \bm s$, $\,\bm{\mathcal{X}}, \bm{\mathcal{L}},\bm\zeta$,
and $\bm{\mathfrak{R}}, \bm{\mathfrak{L}},\bm{\mathfrak{S}}$. Other
fundamental operators in these sets coincide. The three sets are
self-consistent but the operators in these sets have different meanings.
For a charged particle, the first set defines the conventional operators
of the center-of-charge position, the OAM, and the rest-frame spin.
These operators satisfy Eq. (\ref{commutator}). While the NW position
operators in the Dirac representation ($\bm X$) and in the FW one
($\bm x$) do not exactly determine the center-of-mass position of an
ensemble of spinning particles, they define the center-of-charge
position of this
ensemble (see Sec. \ref{Classical}).

The second set characterizes the center-of-mass position, the
\emph{corresponding} OAM, and the laboratory-frame spin.
The quantity $\bm{\mathcal{X}}$ defines the center-of-mass position
of an ensemble of spinning particles. However, the Cartesian
components of the operator $\bm{\mathcal{X}}$ do not commute and
the standard commutation relations for the components of the OAM
and spin are not satisfied either:
\begin{equation}\begin{array}{c}
[\mathcal{X}_i,\mathcal{X}_j]\neq0,\quad [\mathcal{L}_i,\mathcal{L}_j]\neq
ie_{ijk}\mathcal{L}_k,\quad [\zeta_i,\zeta_j]\neq ie_{ijk}\zeta_k,\quad i\neq j;\\
\left[\mathcal{L}_i,\zeta_j\right]\neq0\qquad {\rm for~any~} i,j.
\end{array}\label{commrelcom}
\end{equation}

The third set defines the projected operators and is useful for the
description of the Berry phase effects. For the operators forming
this set, the commutation relations are similar to Eq. (\ref{commrelcom}):
\begin{equation}\begin{array}{c}
[\mathfrak{R}_i,\mathfrak{R}_j]\neq0,\quad [\mathfrak{L}_i,\mathfrak{L}_j]\neq
ie_{ijk}\mathfrak{L}_k,\quad [\mathfrak{S}_i,\mathfrak{S}_j]\neq ie_{ijk}\mathfrak{S}_k,\quad i\neq j;\\
\left[\mathfrak{L}_i,\mathfrak{S}_j\right]\neq0\qquad {\rm for~any~} i,j.
\end{array}\label{commrelprj}
\end{equation}
Certainly, explicit forms of the commutators in Eqs. (\ref{commrelcom})
and (\ref{commrelprj}) differ.

It can be easily shown that the first set is much more convenient than
the second and third sets. Besides a commutative geometry, an important
reason is a definition of electromagnetic and other interactions. It is
very important that the electromagnetic fields act on charges and currents.
Therefore, \emph{the electromagnetic interactions are defined by the
center-of-charge position but not by the center-of-mass one}. The
interaction energy depends on the fields in the center-of-charge point
but not in the center-of-mass one. The same situation takes place for
gravitational, inertial, and weak interactions. In all quantum-mechanical
equations describing the gravitational and inertial interactions (see Refs. \cite{PRD,PRD2,Warszawa,OST,OSTRONG,ostgrav,ostor,ostor2,OSTalkPRD,OSTotal,PRDProca,Honnefscalar}
and references therein), the radius vector relates to the position of the
mass point coinciding with the center of charge for charged particles with
negligible EDMs. Equations of motion obtained with the first (conventional)
set of fundamental operators are fully right while some of these equations
can be noncovariant \cite{PomKh,Obzor}. When the weak interaction is considered
in the framework of QM \cite{ComminsBucksbaum,weakNIM,weakTMP}, the situation is the same.
As a result, there is no reason for a wide use of the second and third sets
of fundamental operators for a description of the fundamental interactions.
In particular, these operators are useless for relativistic quantum chemistry
and physics of heavy atoms. Nevertheless, we agree with Refs. \cite{PomKh,Obzor}
that the noncovariance of equations of motion can be avoided by passing to the
center-of-mass position. For this purpose, the second set of fundamental operators
is useful. Another exception is a determination of the Berry phase effects with
the third set. In other cases, one needs to apply the first set of fundamental
operators.

The necessity of using the mathematical tool of noncommutative geometry significantly
complicates all derivations with the second and third sets. Further, the
laboratory-frame spin $\bm\zeta$, the corresponding OAM $\bm{\mathcal{L}}$, and
the projected operators $\bm{\mathfrak{L}}$ and $\bm{\mathfrak{S}}$ are momentum-dependent.
As a result, the commutation relations for their components given by
Eqs. (\ref{commrelcom}) and (\ref{commrelprj}) are not similar to the commutation
relations (\ref{commutator}) for the components of the total angular momentum $\bm j$.
In addition, the use of the above-mentioned operators prevents one from introducing
the quantum numbers $l$ and $s$ connected with the \emph{conventional}
operators $\bm l$ and $\bm s$. Only the conventional operators belonging to
the first set satisfies the commutation relations similar to those for the
total angular momentum [see Eq. (\ref{commutator})].

The consideration of relativistic operators of the position and spin carried
out in this section leads to the conclusions which agree with the results
obtained in Refs. \cite{SuttorpDeGroot,DeKerfBauerle,PomKh,Ryder,Caban,Bauke,CKT,Obzor}.
However, our conclusions contradict the conclusions which have been made in Refs. \cite{BBBarnett,BB,ReiherWolfBook,electrondensity,BliokhDN,Bliokh2017,BBPRL2017,
QChem1990,QChem1998,SmirnovaBliokh} and are widely used in physics of twisted
electrons and relativistic quantum chemistry. The analysis of the latter
conclusions launched in this section will be finalized in Sec. \ref{selected}.

\section{Related problems of relativistic quantum mechanics}\label{selected}

In this section, we analyze and correct two common errors: a probabilistic
interpretation of a wave function in the Dirac representation and an assertion
about an existence of SOI for a free particle. We also discuss the problem
of \textit{Zitterbewegung}.

\subsection{Probabilistic interpretation of a wave function}\label{probabilistic}

Unfortunately, many scientists
suppose that the Dirac representation corrupting the connection
among energy, momentum, and velocity provides the right distribution
of the probability density and the FW
representation restoring the Schr\"{o}dinger picture of relativistic
QM distorts this density. While this point of view is not correct, it
is presented in almost all papers on twisted (vortex) electrons and,
moreover, prevails in publications devoted to some other problems.
In particular,
the probabilistic interpretation of the wave function in the Dirac
representation is commonly used in relativistic quantum chemistry
(see below). Of course, this
situation is not satisfactory.

It is generally accepted that nonrelativistic Schr\"{o}dinger QM admits
a probabilistic interpretation of the wave function. The classical
center-of-charge position $\bm R$ corresponds to the Schr\"{o}dinger
position operator (the radius vector $\bm x$). In the relativistic case,
the classical center-of-charge position is a counterpart of the FW
position operator which is also equal to the radius vector $\bm x$.
This property has been first established in Ref. \cite{FW} and unambiguously
follows from our analysis. As a result, just the FW wave function being
an expansion of the Schr\"{o}dinger wave function on the relativistic
case admits the probabilistic interpretation. The wave function in the
Dirac representation cannot have such an interpretation \cite{Reply2019}
because the Dirac radius vector $\bm r$ is not the counterpart of the
classical position.

It can also be noted that the components of the Schr\"{o}dinger position
operator commute. Therefore, any quantum-mechanical approach based on a
position operator with \emph{noncommuting} components cannot be a relativistic
extension of the Schr\"{o}dinger QM. In Sec. \ref{Operators}, we have
considered the sets of operators, $\bm{\mathcal{X}}, \bm{\mathcal{L}},\bm\zeta$
and $\bm{\mathfrak{R}}, \bm{\mathfrak{L}},\bm{\mathfrak{S}}$, containing
the laboratory-frame spin $\bm\zeta$ and the projected operators, respectively.
The components of the position operators are noncommuting in the both sets.
As a result, the quantum-mechanical approaches based on these sets lead to
wave functions which cannot be relativistic extensions of the Schr\"{o}dinger
wave functions and cannot have a \emph{direct} probabilistic interpretation.
However, wave functions based on the above-mentioned (second and third) sets
of operators can be derived from the FW wave functions. For any set of fundamental
operators, the classical limit of the FW Hamiltonian coincides with the \emph{corresponding}
classical Hamiltonian. While the Hamiltonians are equal for different sets,
their functional dependencies on the corresponding operators of the position,
OAM, and spin vary. It can be added that the Schr\"{o}dinger-Pauli spin operator
satisfies the commutation relations (\ref{commutator}) which remain valid for
the FW spin operator but are violated for the operators $\bm\zeta$ and
$\bm{\mathfrak{S}}$.

Therefore, the assertion that the quantity
$\varrho_D(\bm r)=\Psi_D^\dag(\bm r)\Psi_D(\bm r)$
is the probability density of the particle position
\cite{BBBarnett,BB,BBPRL2017} is not correct. In fact,
the probability density of the particle position is equal to
$\varrho(\bm x)=\varrho_{FW}(\bm x)=\Psi_{FW}^\dag(\bm x)\Psi_{FW}(\bm x)$
\cite{Reply2019,footnote2}.
This statement has also been made in Refs. \cite{Foldy56,PRDProca} and has been
implicitly used in Refs.
\cite{BarnettPRL,ResonanceTwistedElectrons,PhysRevLettEQM2019,LightArXiv}.
The basic role of the FW representation for a particle in nonstationary
fields has been properly shown in Ref. \cite{ExpectationValue} (see also Sec. \ref{Comparison}).

The quantities $\varrho_D$ and $\varrho_{FW}$ can significantly differ
\cite{BB,Reply2019,BW,Energy3}. A general connection between the Dirac
and FW wave functions at the exact FW transformation has been obtained
in Ref. \cite{Energy3}.  In this case, upper spinors in the two representations
differ only by constant factors and lower FW spinors vanish. An origin of
the difference between $\varrho_D$ and $\varrho_{FW}$ is clear from the
following derivation. Since $\Psi_{FW}=U_{FW}\Psi_{D}$ and $U_{FW}$ is a
self-adjoint unitary operator, the integration of the probability density
results in
$$
\begin{array}{c}
\int{\varrho_{FW}dV}=\int{\Psi_{FW}^\dag\Psi_{FW}dV}
=\int{(\Psi_D^\dag U_{FW}^{-1})(U_{FW}\Psi_D)dV}\\=
\int{\Psi_D^\dag (U_{FW}^{-1}U_{FW}\Psi_D)dV}=\int{\Psi_D^\dag \Psi_DdV}=1,
\end{array}
$$
where the operator $U_{FW}^{-1}$ in $(\Psi_D^\dag U_{FW}^{-1})$ and
$(U_{FW}^{-1}U_{FW}\Psi_D)$ acts to the left and to the right, respectively.
However, the self-adjointness of operators manifests at the integration
but cannot be used in any fixed point of a domain of definition. Therefore,
$$
\Psi_{FW}^\dag\Psi_{FW}
=(\Psi_D^\dag U_{FW}^{-1})(U_{FW}\Psi_D)\neq \Psi_D^\dag \Psi_D
$$
and $\varrho_{FW}\neq\varrho_D$.

The probabilistic interpretation of the FW wave function allows one to
calculate expectation values of all operators. In particular, the mean
squared radius $<r^2>$ and the quadrupole moment tensor $Q_{ij}$ are given by
\begin{equation}
<r^2>=\int{\Psi_{FW}^\dag\bm x^2\Psi_{FW}dV},\qquad Q_{ij}
=\int{\Psi_{FW}^\dag(3x_ix_j-\bm x^2\delta_{ij})\Psi_{FW}dV}.
\label{COMOAMLRT}
\end{equation}

In relativistic quantum chemistry, the term ``FW transformation'' is used
for the original transformation by Foldy and Wouthuysen \cite{FW} and the
relativistic FW transformation is called the ``Douglas-Kroll-Hess transformation''.
The latter transformation can be carried out with any needed accuracy.
For this purpose, analytical or numerical calculations can be fulfilled.
In contemporary relativistic quantum chemistry, the point of view contradictory
to our analysis is generally accepted (see Sec. 15.2 in Ref. \cite{ReiherWolfBook}
and Refs. \cite{electrondensity,QChem1990,QChem1998}). It is supposed that
expectation values of operators are defined in the \emph{Dirac} representation.
In this case, the use of the FW representation needs the transformation of
operators to the FW representation and expectation values of \emph{transformed}
operators are determined. The expectation values of any operator $A$ in the
Dirac and FW representations are defined by
\begin{equation}
\begin{array}{c}
<A>\equiv<A_D>=\int{\Psi_{D}^\dag A\Psi_{D}dV}
=\int{\Psi_{FW}^\dag (U_{FW}AU_{FW}^{-1})\Psi_{FW}dV}
=\int{\Psi_{FW}^\dag A'\Psi_{FW}dV},\\
\tilde{A}\equiv<A_{FW}>=\int{\Psi_{FW}^\dag A\Psi_{FW}dV},\qquad A'=U_{FW}AU_{FW}^{-1}.
\end{array}
\label{anyoA}
\end{equation}
The difference
\begin{equation}
PCE(A) = <A> -\tilde{A}
\label{PCE}
\end{equation}
is called the ``picture change error'' \cite{ReiherWolfBook,electrondensity,QChem1990,QChem1998}.
For example, the formula used for a calculation of the quadrupole moment tensor reads
\begin{equation}
Q_{ij}=\int{\Psi_{D}^\dag(3x_ix_j-\bm r^2\delta_{ij})\Psi_{D}dV}.
\label{rqcQM}
\end{equation}
Our analysis unambiguously shows that this definition is not correct. The picture
change error is, indeed, equal to zero and the expectation values of all
operators should be defined in the FW representation. Therefore, all results
obtained in relativistic quantum chemistry with Eq. (\ref{PCE}) should be reconsidered.

\subsection{Spin-orbit interaction for a free particle}\label{spinorbitcoupling}

The analysis presented in Secs. \ref{Classical} and \ref{Comparison} shows the
correspondence between the classical
rest-frame spin $\bm S$, the Pauli spin, and the FW mean-spin operator
($\bm s=\hbar\bm\Sigma/2$ in the FW
representation). This correspondence has been discovered by Foldy and Wouthuysen
\cite{FW} and has been
shown
in Refs. \cite{JordanMuku,AcharyaSudarshan,Wightman,FGursey,Fronsdal,Bacry,OConnell,
Kalnayetal,Foldy,BGS,Mattews,Costella}. Evidently, the FW mean-spin operator
commutes with the FW Hamiltonian (\ref{HamFW}) and the FW mean-OAM operator
($\bm l=\bm x\times\bm p$ in the FW representation). Therefore, the SOI cannot
exist for the conventional rest-frame spin operator $\bm s$ and the corresponding
OAM operator $\bm l$. In the Dirac representation, the operator $\hbar\bm\Sigma/2$
does not commute with the Dirac Hamiltonian. However, it does not describe the
conventional spin defined by Eq. (\ref{menspin}). Thus, applying the first set
of fundamental operators (see Sec. \ref{Operators}) leads to the nonexistence of the SOI.

In Refs. \cite{BliokhDN,Bliokh2017,SmirnovaBliokh}, the existence of the SOI for a
free Dirac particle has been claimed. This statement is based on the assumption that
the quantum-mechanical counterparts of the position, OAM, spin, and other fundamental
classical variables are the corresponding operators in the \emph{Dirac} representation.
However, it has been explicitly shown in numerous publications considered in detail in
Secs. \ref{Classical}--\ref{Operators} that these counterparts are the corresponding
operators in the FW representation, $\bm x, \bm l, \bm s$. In relation to the spin,
it has been made in Refs. \cite{FW,FG,FGursey,BGS,Mattews,Costella,Ryder,Caban,Bauke}.
The connection between the classical and quantum-mechanical descriptions of the spin
has been expounded in Secs. \ref{Classical}--\ref{Operators}. All these results contradict
to the key statement in Ref. \cite{Bliokh2017} that the spin is defined by the
operator $\bm s_D=\hbar\bm\Sigma/2$ in the \emph{Dirac} representation.

It is instructive to discuss why this statement leads to the SOI for a free particle.
The use of the projected operators analyzed in Sec. \ref{Operators} considerably
simplifies the consideration of the SOI.
There is some similarity between operators in the Dirac representation and the
corresponding projected operators. For any localized particle state, they have the
same expectation values \cite{Thaller,Bliokh2017}. In particular, the spin operator
in the \emph{Dirac} representation $\bm s_D$ transformed to the FW representation
takes the form \cite{FW}
\begin{equation}
\bm{\mathfrak{s}}_{D\rightarrow FW}=\frac{m\bm\Sigma}{2\epsilon}
+\frac{\bm p(\bm p\cdot\bm\Sigma)}{2\epsilon(\epsilon+m)}
+\frac{i\bm \gamma\times\bm p}{2\epsilon}.
\label{DraFW}
\end{equation}
Its similarity to the projected spin operator in the FW representation
$\bm{\mathfrak{S}}_{FW}$ defined by Eq. (\ref{poDis}) is evident. Since the
lower FW spinor for positive-energy states and the upper FW spinor for
negative-energy ones are equal to zero, the expectation values of the operators
$\bm{\mathfrak{S}}_{FW}$ and $\bm{\mathfrak{s}}_{D\rightarrow FW}$ coincide.
Certainly, they also coincide with the expectation values of these operators
in the Dirac representation, $<\bm{\mathfrak{S}}_{D}>$ and $<\bm{s}_D>$.

It is worth mentioning that the Dirac operators and the corresponding projected
ones are not equivalent. In particular, the Dirac spin operator satisfies the
standard algebra $\left[(s_D)_i,(s_D)_j\right]=ie_{ijk}(s_D)_k$ $(i,j,k=1,2,3)$.
Its square is equal to $\bm s_D^2=s(s+1)\mathcal{I}=3\mathcal{I}/4$, where
$\mathcal{I}$ is the $2\times2$ unit matrix. The Dirac position operator has
commutative components $\left([r_i,r_j]=0\right)$. These properties remain
valid in any representation. The corresponding projected operators do not
satisfy these properties [see Eq. (\ref{commrelprj})].
Because of the nonequivalence of the Dirac and projected operators and the
inconsistency of the former operators with any fundamental classical variables,
the Dirac operators are useless. In contrast to them, the projected operators
may be useful for a solution of some physical problems. However, their
application needs noncommutative geometry (see Sec. \ref{Operators}).

The connection between the expectation values of the Dirac spin operator
and the rest-frame spin (e.g., ``mean spin angular momentum'' \cite{FW})
follows from Eqs. (\ref{poDis}) and (\ref{DraFW}) and is given by Eq. (7)
in Ref. \cite{Bliokh2017}:
\begin{equation}
<\bm{s}_{D}>=\frac{m}{\epsilon}<\bm{s}>
+\frac{\bm p(\bm p\cdot<\bm{s}>)}{\epsilon(\epsilon+m)}.
\label{DiracBliokh}
\end{equation}

In disagreement with Ref. \cite{Bliokh2017}, the Dirac spin operator and the
projected spin substantially differ from the spatial parts of the four-component
spin operator and the Pauli-Lubanski four-vector, $\bm a$ and $\bm W=m\bm a$,
respectively. This fact has already been mentioned in Sec. \ref{Operators}.

The opposite statement presented in Ref. \cite{Bliokh2017} is rather strange
because Eqs. (15) and (16) in this paper give the correct relation between the
projected and Pauli-Lubanski operators, $\bm{\mathfrak{S}}_{FW}=\bm W/\epsilon$
[see also Eq. (\ref{PLsFW}) in Sec. \ref{Operators}].

It is also claimed in Ref. \cite{Bliokh2017} that the projected operators are
covariant. However, the results
obtained in Refs. \cite{SuttorpDeGroot,DeKerfBauerle,KhPom,PomKh,Obzor}
unambiguously show that covariant
equations of motion can be obtained \emph{only} with the laboratory-frame
spin $\bm\zeta$ and the corresponding
operators of the position and OAM, $\bm{\mathcal{X}}$ and $\bm{\mathcal{L}}$.
These quantities form the
\emph{second} set of fundamental operators (see Sec. \ref{Operators}). As a
result, the use of the projected operators of the position, OAM, and spin
leads to noncovariant equations of motion for the spin and momentum in
the presence of external fields. Even if some equations of spin motion are
covariant, it is not so in the
general case. This conclusion remains valid for the corresponding operators
in the Dirac representation, $\bm r,\bm l_D,\bm s_D$. Moreover, even the application of
$\bm a$ as \emph{a spin part of the total angular momentum} $\bm j$ with the
corresponding redefinition of the
position operator cannot result in covariant equations of motion.
In this case, $\bm j=\bm\varrho\times\bm p+\bm a$, where $\bm\varrho$ is the
corresponding position operator. The quantities $\bm a$ and $\bm j$ are
covariant. However, $\bm a$ is a spatial part of a four-vector and $\bm j$
is formed by spatial components of an antisymmetric tensor. Therefore, the
quantities $\bm a$ and $\bm j$ are dissimilar and their simultaneous use
does not lead to the covariant equations of motion.
The problem of the covariant fundamental operators and the covariant equations
of motion has been definitively solved in Refs. \cite{SuttorpDeGroot,DeKerfBauerle}.
However, these papers
are not cited in Ref. \cite{Bliokh2017}.

We should also mention that Eq. (7) in Ref. \cite{Bliokh2017} coinciding with
Eq. (\ref{DiracBliokh}) in the present study has nothing in common with
``the Lorentz boost to the rest frame''. It describes the connection between
the expectation values of the Dirac spin $<\bm s_D>$ (defined as $<\bm S>$ in
Ref. \cite{Bliokh2017}) and the FW (rest-frame) spin $<\bm s>$. The Lorentz
boosts define only the connections between the expectation values of the FW
spin and the spatial parts of either the four-component spin vector $(<\bm a>)$
or the antisymmetric spin tensor $(<\bm\zeta>)$. These connections are given by
\begin{equation}
<\bm a>=<\bm s>+\frac{\bm p(\bm p\cdot<\bm{s}>)}{m(\epsilon+m)},
\qquad <\bm\zeta>=\frac\epsilon m<\bm s>-\frac{\bm p(\bm p\cdot<\bm{s}>)}{m(\epsilon+m)}.
\label{fourspina}
\end{equation}
Evidently, they substantially differ from the connection described by Eq.
(\ref{DiracBliokh}) (Eq. (7) in Ref. \cite{Bliokh2017}).

Our next 
comment relates to one of principal Pryce's assumptions \cite{Pryce} that a
choice of the fundamental variables
$\bm Q, \bm {L},\bm{S}$ is not unique and the use of different sets of these
variables is possible. The revision
of this assumption made in Ref. \cite{Bliokh2017} is unsubstantiated. The
authors of this paper have considered the
first and third sets of fundamental operators and have concluded that only
the latter set of these operators
correctly and exhaustively describes physical phenomena. However, it is
clear from Ref. \cite{Pryce} and subsequent publications that the use of
the ten-parameter Poincar\'{e} group leaves a room for different definitions
of the position and spin operators satisfying Eqs. (\ref{spinOAM}) and (\ref{tomFW}).
Of course, useful definitions should have appropriate substantiations.
Nevertheless, \emph{all} well-substantiated position, OAM, and spin operators
are correct. We have analyzed the three sets of fundamental operators. Each
of them is useful for a description of some physical phenomena and dynamical
equations for the spin and momentum obtained with all the sets should agree.
Furthermore, any other set of fundamental operators (even based on a confusion)
leads to correct dynamical equations. Certainly, different sets are not equally
convenient. We mentioned above that the second and third sets of fundamental
operators are not useful for relativistic quantum chemistry and physics of heavy atoms.
The preferences of the first set are the commutative geometry, simple commutation
relations and the independence of potentials and strengths of external fields from
the momentum and spin of the particle. These preferences are seen from the following
example. When the first set is used, the scalar potential of an electric field in
the FW representation has the simple form $\phi(\bm x)$. For the set of projected
operators, it is given by
$\phi\left(\bm{\mathfrak{R}}_{FW}+\frac{\bm{\mathfrak{S}}_{FW}\times\bm p}{m(\epsilon+m)}\right)$.
While the potential $\phi\left(\bm{\mathfrak{R}}_{FW}\right)$ incorporates the SOI
with an external electric field \cite{Bliokh2017}, the need for use of noncommutative
geometry makes this set to be inconvenient.

The analysis fulfilled unambiguously shows that there is not the SOI if the terms
``spin'' and ``OAM'' denote the \emph{conventional} rest-frame spin operator $\bm s$
and the corresponding OAM operator $\bm l$. Since different definitions of the
fundamental operators of the position, OAM, and spin are admissible, the existence
or nonexistence of the SOI for free particles depends on these definitions. More
precisely, the SOI does not exist for the conventional fundamental operators
containing the first set but exists for the operators forming other sets. However,
it would be misleading to assert that the SOI exists but to omit the specification
that this effect takes place, e.g., for projected operators based on noncommutative
geometry and specific commutation relations. We also underline that the fundamental
classical variables correspond to the related fundamental operators in the FW
representation while the Dirac representation distorts these operators.

We should also comment the statement \cite{Bliokh2017} that the FW representation
``cannot be used for massless particles''. The application of relativistic methods
of the FW transformation explicitly shows the inaccuracy of this statement. Even
the early paper by Blount \cite{Blount} has been demonstrated (despite some imperfections)
the validity of the FW
representation in the massless limit $m\rightarrow0$. The appropriateness of the
FW representation in the massless limit is also clearly seen
from Eqs. (\ref{FWHamEDM})-(\ref{EDMeq12}), (\ref{eqFWEDM12}), (\ref{Hamgnif}),
(\ref{eqFWsca}), (\ref{eqFWEDMunit}) and (\ref{TrHamfnEEV}) \cite{footnote3}.
The only important difference between the FW transformations for massive and massless
particles is the loss of classical interpretation of the spin operator
$\bm s=\hbar\bm\Sigma/2$ when $m=0$. It can be added that the FW representation
has been used for the photon in Refs. \cite{LightArXiv,Barnett-FWQM}.

Nevertheless, we should note that particle physics does not support the smooth
transition to the massless limit. Spin-0 and spin-1/2 particles can be considered
as exceptions. For the photon, the FW transformation significantly differs from
that for massive spin-1 particles \cite{PhysRevDspinunit}.
It is well known that massive spin-1 particles can have the helicity $0,\pm1$ but
the photon cannot exist in a helicity zero state. The rest-frame spin of massive
spin-1 particles is described by the $3\times3$ matrix (\ref{eqspinunit}). Its
use for a fixed momentum direction allows one to obtain the three above-mentioned
eigenvalues of the helicity operator. However, this matrix cannot reproduce the
two helicity eigenvalues $\pm1$ of the photon. While the matrix (\ref{eqspinunit})
is also applied for the photon \cite{Barnett-FWQM,Dirac-like,BialynickiBirulaPhysScr},
a possibility to connect it with the photon spin seems to be, at least, doubtful.
This simple analysis shows a deep difference between massless particles with the
spins 1/2 and 1. Therefore, results obtained for the photon cannot be directly
applied to a massless spin-1/2 particle. For the photon, the laboratory-frame
spin operator leading to the SOI can be used.

We should also note that QM cannot provide for an exhaustive description of
massless particles. For this purpose,  the mathematical tool of quantum field
theory is needed. Nevertheless, some specific properties of massless particles
can be studied in the framework of QM. In this case, the FW transformation can
be helpful even for the photon (see Refs. \cite{Barnett-FWQM,LightArXiv}).

\subsection{\textit{Zitterbewegung}}\label{ZitterbewegungQM}

\textit{Zitterbewegung} is a well-known effect consisting of a superfast trembling
motion of a free Dirac particle first described by Schr\"{o}dinger \cite{ZitterbewegungSc}.
This effect is also known for a scalar particle \cite{ZitterbewegungGG,ZitterbewegungFF}.
Our preceding analysis perfectly agrees with the conclusions about the origin and
observability
of this effect made in Refs. \cite{ZitterbewegungFF,Zitterbewegungbook,ZitterbewegungOC}.

As is well known, the Dirac velocity operator is given by
\begin{equation}
\bm v_D\equiv\frac{d\bm r}{dt}=i[{\cal H}_D,\bm r]=\bm\alpha.
\label{Diracvlct}
\end{equation}
This operator is time-dependent:
\begin{equation}
\frac{d\bm\alpha}{dt}=i[{\cal H}_D,\bm\alpha]
=i\{\bm\alpha,{\cal H}_D\}-2i\bm\alpha{\cal H}_D
=2i(\bm p-\bm\alpha{\cal H}_D).
\label{Diraccel}
\end{equation}

The problem is usually considered in the Heisenberg picture:
\begin{equation}
\bm v_D(t)=e^{i{\cal H}_Dt}\bm\alpha e^{-i{\cal H}_Dt}.
\label{Heispte}
\end{equation}
In the Schr\"{o}dinger picture, the result is the same.
We suppose that
the eigenvalues of the momentum and Hamiltonian operators are
$\bm{\mathfrak{p}}$ and $H$, respectively. In this case,
Eq. (\ref{Diraccel}) can be presented in terms of the Dirac velocity operator:
\begin{equation}
\frac{d\bm v_D}{dt}=2i(\bm{\mathfrak{p}}-\bm v_D H).
\label{Dprvelo}
\end{equation}
Its integration shows the oscillatory behavior of the Dirac velocity:
\begin{equation}
\bm v_D(t)=\left[\bm v_D(0)-\frac{\bm{\mathfrak{p}}}{H}\right]e^{-2iHt}
+\frac{\bm{\mathfrak{p}}}{H}.
\label{Dirvele}
\end{equation}
The evolution of the Dirac position operator is given by
\begin{equation}
\bm r_D(t)=\bm r_D(0)+\frac{\bm{\mathfrak{p}}t}{H}+\frac{i}{2H}\left[\bm v_D(0)
-\frac{\bm{\mathfrak{p}}}{H}\right]\left(e^{-2iHt}-1\right).
\label{Dirpoev}
\end{equation}

For a free scalar (spin-0) particle, the initial Feshbach-Villars Hamiltonian reads \cite{FV}
\begin{equation}
{\cal H}_{FV}=\rho_3m+\left(\rho_3+i\rho_2\right)\frac{\bm p^2}{2m}.
\label{HamFV}
\end{equation}
The velocity operator in the Feshbach-Villars representation is equal to
\begin{equation}
\bm v_{FV}=\left(\rho_3+i\rho_2\right)\frac{\bm p}{m}.
\label{velocFV}
\end{equation}
The corresponding acceleration operator is defined by the equation similar to Eq.
(\ref{Diraccel}) \cite{ZitterbewegungFF}:
\begin{equation}
\frac{d\bm v_{FV}}{dt}=i[{\cal H}_{FV},\bm v_{FV}]=i\{\bm v_{FV},{\cal H}_{FV}\}
-2i\bm v_{FV}{\cal H}_{FV}=2i(\bm p-\bm v_{FV}{\cal H}_{FV}).
\label{FVaccel}
\end{equation} We suppose that
the eigenvalues of the momentum and Hamiltonian operators are $\bm{\mathfrak{p}}$
and $H$, respectively.
As a result, the final equations of dynamics of the free scalar particle
\cite{ZitterbewegungFF} are equivalent to the corresponding equations for the
Dirac particle:
\begin{equation}
\bm v_{FV}(t)=\left[\bm v_{FV}(0)-\frac{\bm{\mathfrak{p}}}{H}\right]e^{-2iHt}
+\frac{\bm{\mathfrak{p}}}{H},
\label{FVe}
\end{equation}
\begin{equation}
\bm r_{FV}(t)=\bm r_{FV}(0)+\frac{\bm{\mathfrak{p}}t}{H}+\frac{i}{2H}\left[\bm v_{FV}(0)
-\frac{\bm{\mathfrak{p}}}{H}\right]\left(e^{-2iHt}-1\right).
\label{FVpoe}
\end{equation}

However, it has been pointed out \cite{OConnell} that the operators $\bm p$ and
$\bm v$ can be proportional for free particles with any spin. The proportionality
of these operators vanishing the acceleration ($d\bm v/(dt)=0$) can be achieved
by the FW transformation. In the FW representation, the Dirac Hamiltonian takes
the form (\ref{HamFW}) and the velocity operator is given by
\begin{equation}
\bm v_{FW}=\beta\frac{\bm p}{\sqrt{m^2+\bm p^2}}=\frac{\bm p}{{\cal H}_{FW}}.
\label{eqvelmm}
\end{equation} Similar relations can be obtained for particles with any spin.

It has been shown in Ref. \cite{ZitterbewegungFF} that \textit{Zitterbewegung}
is the result of the interference between positive- and negative-energy states.
It disappears for the ``mean position operator'' \cite{FW}, the position operator
in the FW representation \cite{ZitterbewegungFF,Zitterbewegungbook,ZitterbewegungOC}.
``\textit{Zitterbewegung} was found to be a feature of a particular choice of coordinate
operator associated with Dirac's formulation of relativistic electron theory''
(see Ref. \cite{Zitterbewegungbook}, p. 334). It can be removed by carrying out
the unitary transformation to the FW representation. Experiments do not distinguish
between equally valid but different representations leading to the same observables
and the transition to the FW representation does not change the physics \cite{ZitterbewegungOC}.
It can be concluded that\textit{ Zitterbewegung} is not an observable \cite{ZitterbewegungOC}.

Our analysis fully agrees with this conclusion. The derivations presented in
this subsection show that \textit{Zitterbewegung }is an effect attributed to
the Dirac and Feshbach-Villars position and velocity operators but not to the
corresponding FW operators. However, just the FW position and velocity operators
are the quantum-mechanical counterparts of the classical position and velocity.
In the Dirac representation, these quantum-mechanical counterparts are defined
by the operators $\bm X$ [see Eq. (\ref{meanpos})] and $d\bm X/(dt)$. For the
latter operators, \textit{Zitterbewegung} does not take place in any representation.
Our analysis shows that the Dirac and Feshbach-Villars position and velocity
operators are not the quantum-mechanical counterparts of the classical position
and velocity and, in accordance with Ref. \cite{ZitterbewegungOC}, \textit{Zitterbewegung}
cannot be observed.

However, there exists an effect which is more or less similar to
\textit{Zitterbewegung}. Feshbach and Villars \cite{FV} have obtained
the eigenfunctions of the mean position operator $\bm X$ and have proven
that they are not localized in the configuration space but are extended
over a radius of the order of the Compton wavelength. These eigenstates
are the narrowest possible free wave packets composed only of positive
energy states whose behavior agrees with the nonrelativistic (Schr\"{o}dinger)
pattern. The nonlocality of the particle position takes also place for spinning
particles \cite{Zitterbewegungbook,Sakurai}. It has been emphasized by
Sakurai \cite{Sakurai} that ``the nonlocality of $\bm X$ is the price we
must pay'' for the absence of \textit{Zitterbewegung}. This indirect connection
between the nonlocality of the particle position and \textit{Zitterbewegung}
has also been considered in other works (see, e.g., Refs.
\cite{ZitterbewegungRo,ZitterbewegungBD,ZitterbewegungCapelle}).
The nonlocality of the particle position manifests in the Darwin interaction
defined by Eq. (\ref{eq33new}).

\section{Discussion and summary}\label{Discussions}

The goal of the present study is a change of the paradoxical contemporary
situation in the relativistic QM when the forms of the position and spin operators
securely established sixty years ago are ``forgotten'' while
incorrect and unsubstantiated definitions of these operators are widely used.
The Dirac representation distorts the connection among the energy, momentum,
and velocity operators. Therefore, it is too optimistic to believe that the
Dirac operators of the position and spin, $\bm r$ and $\bm s_D$, are relativistic
extensions of the corresponding Schr\"{o}dinger-Pauli operators and quantum-mechanical
counterparts of the classical position and spin. Indeed, there are not any serious
arguments in favor of this point of view which has been proclaimed in Refs. \cite{BBBarnett,BB,ReiherWolfBook,electrondensity,QChem1990,QChem1998,BBPRL2017}
and has been explicitly or implicitly presented in almost all papers devoted to
twisted electrons.
Some exceptions, in particular, Refs. \cite{BarnettPRL,ResonanceTwistedElectrons,
PhysRevLettEQM2019,LightArXiv} are not numerous.
Paradoxically, the correct results obtained for the position and spin operators
sixty years ago were widely discussed \cite{FW,FG,CurrieRevModPhys,JordanMuku,
BakamjianThomas,Foldy56,Foldy61,Wightman,
FGursey,Fronsdal,Bacry,OConnell,Kalnayetal,Foldy,BGS,Mattews,Costella,Ryder,Caban,Bauke}.
Nevertheless, the researchers holding the opposite point of view never carefully
considered the arguments obtained in the above-mentioned publications in favor of
a definition of the fundamental operators in the FW representation. The proof of
this definition carried out in Refs. \cite{FW,FG,CurrieRevModPhys,JordanMuku,
BakamjianThomas,Foldy56,Foldy61,Wightman,
FGursey,Fronsdal,Bacry,OConnell,Kalnayetal,Foldy,BGS,Mattews,Costella,Ryder,Caban,Bauke}
is straightforward (see Sec. \ref{Classical}).

More recent achievements allowing one to perform the FW transformation for
relativistic scalar and spinning particles in external fields (see Sec. \ref{Relativistic})
have allowed us to give new important arguments in support of this definition.
We have shown in Sec. \ref{Comparison} that the position and spin operators
corresponding to the conventional classical variables (the radius vector $\bm R$
and the rest-frame spin $\bm S$) are the FW operators $\bm x$ and $\bm s=\hbar\bm\Sigma/2$
and their transforms to other representations, including the Dirac representation.
This property remains valid in the presence of external fields. Our analysis
unambiguously shows that the fundamental spin operator is defined in the particle
rest frame but not in the instantaneously accompanying one.

The main result of the present study is the comparative analysis of alternative
definitions of the position and spin operators.
Certainly, the use of the Dirac position operator $\bm r$ as a quantum-mechanical
counterpart of the classical position variable brings confusion. A determination
of the probability density with Dirac eigenfunctions \cite{BBBarnett,BB,BBPRL2017}
distorts the electron charge distribution in a free space and atoms. In relativistic
quantum chemistry, a calculation of expectation values with FW eigenfunctions but
with FW transforms of \emph{Dirac} operators
\cite{ReiherWolfBook,electrondensity,QChem1990,QChem1998} leads to incorrect
results due to unnecessary corrections for the picture change errors.

A calculation of expectation values of the spin with the Dirac spin operator
$\bm s_D$ \cite{BliokhDN,Bliokh2017,SmirnovaBliokh} has similar consequences.
As a result, the illusory effect of the SOI appears. The SOI does not exist
for a free particle if the terms ``spin'' and ``OAM'' define the \emph{conventional}
spin and OAM operators which, in particular, satisfy the commutation relations
$\left[s_i,s_j\right]=ie_{ijk}s_k,\,\left[l_i,l_j\right]=ie_{ijk}l_k,\, \left[l_i,s_j\right]=0$.
Other definitions of these operators are possible, but all of them do not
satisfy these commutation relations and are based on noncommutative geometry.
For such definitions, the spin-orbit interaction can exist. The Dirac spin
operator has the same expectation values as the projected spin operator. Both
of them substantially differ from the two covariant spin operators given by
the spatial parts of the four-component spin vector and the antisymmetric spin
tensor, $\bm a$ and $\bm\zeta$, respectively. Unfortunately, Refs.
\cite{BliokhDN,Bliokh2017,SmirnovaBliokh} present the misleading conclusion about
the SOI. The problem of the SOI is a matter of a definition of the spin. The
conventional QM based on commutative geometry leads to the nonexistence of the SOI.

We can conclude that the basic representation in relativistic QM is the FW one
because it provides for a direct similarity between the relativistic
quantum-mechanical operators and the classical variables.

\begin{acknowledgments}
This work was supported by the Belarusian Republican Foundation
for Fundamental Research
(Grant No. $\Phi$18D-002), by the National Natural Science
Foundation of China (Grants No. 11975320 and No. 11805242),
and by the the Chinese Academy of Sciences
President's International Fellowship Initiative (No. 2019VMA0019).
A. J. S. also acknowledges hospitality and support by the
Institute of Modern
Physics of the Chinese Academy of Sciences. The authors are grateful to A. Deriglazov for paying their attention to Refs. \cite{Deriglazov1,Deriglazov2}.
\end{acknowledgments}


\begin{thebibliography}{}
\bibitem{footnote2}
In expressions for $\varrho_{D}$ and $\varrho_{FW}$, the variables $\bm r$ and $\bm x$ are identical.
\bibitem{footnote3}
In the massless limit, one should use the quantities $\mu_0,\,\mu'$, and $d$
and should not pass to the quantities $a$ and $\eta$ defined in the text.
\end{thebibliography}
\end{document}